\documentclass[pra,aps,twocolumn,floatfix,superscriptaddress,showpacs,A4]{revtex4-1}
\usepackage{graphicx,graphics,psfrag,amsmath,amsfonts,calc,epsfig,color,mathtools,bm}

\begin{document}

\title{Quantum phase transitions and Berezinskii-Kosterlitz-Thouless temperature \\ in a two-dimensional spin-orbit-coupled Fermi gas}
	
\author{Jeroen P.A. Devreese}
\address{TQC, Universiteit Antwerpen, B-2610 Antwerpen, Belgium}
\affiliation{School of Physics, Georgia Institute of Technology, 
Atlanta, 30332, USA}

\author{Jacques Tempere}
\address{TQC, Universiteit Antwerpen, B-2610 Antwerpen, Belgium}
\affiliation{Lyman Laboratory of Physics, Harvard University, Cambridge, Massachusetts 02138, USA}

\author{Carlos A.R. S\'a de Melo}
\address{School of Physics, Georgia Institute of Technology, 
Atlanta, 30332, USA}

\date{\today}

\begin{abstract}
We study the effect of spin-orbit coupling on both the zero-temperature and non-zero temperature behavior of a two-dimensional (2D) Fermi gas. We include a generic combination of Rashba and Dresselhaus terms into the system Hamiltonian, which allows us to study both the experimentally relevant equal-Rashba-Dresselhaus (ERD) limit and the Rashba-only (RO) limit. At zero temperature, we derive the phase diagram as a function of the two-body binding energy and Zeeman field. In the ERD case, this phase diagram reveals several topologically distinct uniform superfluid phases, classified according to the nodal structure of the quasiparticle excitation energies. Furthermore, we use a momentum dependent SU(2)-rotation to transform the system into a generalized helicity basis, revealing that spin-orbit coupling induces a triplet pairing component of the order parameter. At non-zero temperature, we study the Berezinskii-Kosterlitz-Thouless (BKT) phase transition by including phase fluctuations of the order parameter up to second order. We show that the superfluid density becomes anisotropic due to the presence of spin-orbit coupling (except in the RO case). This leads both to elliptic vortices and antivortices, and to anisotropic sound velocities. The latter prove to be sensitive to quantum phase transitions between topologically distinct phases. We show further that at a fixed non-zero Zeeman field, the BKT critical temperature is increased by the presence of ERD spin-orbit coupling. Subsequently, we demonstrate that the Clogston limit becomes infinite: $T_{\rm{BKT}}$ remains non-zero at all finite values of the Zeeman field. We conclude by extending the quantum phase transition lines to non-zero temperature, using the nodal structure of the quasiparticle spectrum, thus connecting the BKT critical temperature with the zero-temperature results.
\end{abstract}
\maketitle
\section{Introduction}
Spin-orbit coupling, the interaction of a particle's spin with its motion, is an essential ingredient in many quantum mechanical phenomena. In atomic physics, this effect arises from the interaction between the electron's magnetic moment and the magnetic field generated by the electron's orbital motion, giving rise to the fine-structure splitting. In condensed matter physics, spin-orbit coupling leads to intriguing phenomena such as topological insulators \cite{Hasan Kane RMP}, the quantum spin-Hall effect \cite{Spin Quantum Hall Sinova/MacDonald,Spin Quantum Hall Kato/Awschalom} and Weyl fermions \cite{Weyl fermions}. However, in these cases, the strength of the spin-orbit coupling is intrinsic and, moreover, the complex structure of the materials used is not always known, making theoretical modeling an arduous task. \newline \indent
By contrast, ultracold atomic gases offer a versatile system in which parameters such as the interaction strength, the spin-imbalance, the dimensionality and geometry can be freely adjusted \cite{Ultracold Gases}, making them ideally suited for quantum simulation of many-body systems. However, because the atoms used in ultracold gases are neutral, creating artificial spin-orbit coupling required the exploration of new techniques. More specifically, the use of two-photon Raman transitions was suggested theoretically \cite{Rashba SOC theoretical proposal 1,Rashba SOC theoretical proposal 2,Rashba SOC theoretical proposal 3} and shortly thereafter implemented for bosonic gases \cite{SOC Bosons}. Subsequently, spin-orbit coupling was created in systems of non-interacting fermions \cite{SOC Fermions 1, SOC Fermions 2}. Recently, the interacting spin-orbit-coupled Fermi gas near a Feshbach resonance has also been realized \cite{SOC interacting Fermions} and the formation of Feshbach molecules was investigated theoretically \cite{Doga Carlos arXiv}. The type of spin-orbit coupling achieved in these systems is that of equal Rashba \cite{Rashba SOC} and Dresselhaus \cite{Dresselhaus SOC} strength (ERD), which up till now is the only form realized experimentally.\newline \indent
These seminal experiments have sparked a wide range of suggestions for new experimental set-ups. Proposals to create spin-orbit coupling without the use of Raman dressing (which suffers from heating problems) include rf dressing with an atom chip \cite{Atom Chip} and using ladder-like optical lattices \cite{Ladder-like optical lattices}. Furthermore, many proposals have emerged for the creation of Rashba-only spin-orbit coupling, including the creation of degenerate dark states using tripod laser coupling \cite{Dark States 1, Dark States 2} and generalizing the Raman scheme used in the aforementioned experiments \cite{Generalized Raman 1, Generalized Raman 2}. For a more complete overview of the experimental achievements in this rapidly developing field, we refer to the following excellent review papers \cite{Dalibard review paper, Spielman review paper, Zhai review paper}.\newline \indent
The first theoretical studies of spin-orbit coupling in ultracold gases focused on the three-dimensional (3D) case, with either Rashba-only (RO) coupling \cite{3D RO SOC 1,3D RO SOC 2,3D RO SOC 3,3D RO SOC 4} or ERD coupling \cite{3D ERD SOC 1,3D ERD SOC 2}. Recently, the two-dimensional (2D) RO case has also received wide attention \cite{2D RO case 1,2D RO case 2,2D RO case 3}, as well as the 2D ERD case \cite{Li Carlos arXiv}, in part due to the experimental creation of a 2D interacting Fermi gas \cite{2D Fermi gas 1,2D Fermi gas 2,2D Fermi gas 3,2D Fermi gas 4} and by its relation to topological superfluids \cite{Topological superfluid 1,Topological superfluid 2}. However, in 2D, at non-zero temperature, a phase transition from a quasi-condensate to a non-superfluid paired phase arises due to the Berezinskii-Kosterlitz-Thouless (BKT) mechanism, which involves the unbinding of vortex-antivortex pairs \cite{BKT 1,BKT 2}. To capture the physics of this phenomenon it is essential to go beyond the saddle-point (mean-field) approximation and include fluctuations of the phase of the order parameter into the description \cite{phase fluctuations 1,phase fluctuations 2,phase fluctuations 3}. We performed this calculation for the 2D case with generic spin-orbit coupling, which was reported in a recent Letter \cite{PRL Jeroen Jacques Carlos}. In the current paper, we will discuss the full mathematical details of the aforementioned calculation, as well as present additional results for the zero-temperature case and for the BKT critical temperature.\newline \indent
The remainder of this paper is divided in two parts: saddle point and fluctuations. In Sec. \ref{Section II}, we develop and discuss the saddle-point case. We start in Sec. \ref{Section II A} by introducing the system Hamiltonian and setting up the functional-integral formalism, which we use throughout the paper. A derivation of the saddle-point thermodynamic potential is shown in Sec. \ref{Section II B}. Subsequently, in Sec. \ref{Section II C}, we make a momentum dependent transformation to the generalized helicity basis, which shows the emergence of a triplet component of the order parameter. In Sec. \ref{Section II D}, we define the topologically distinct uniform superfluid phases of the system, based on the nodal structure of the quasiparticle excitation spectra. Finally, in Sec. \ref{Section II E}, we calculate the zero-temperature phase diagram as a function of the two-body binding energy and Zeeman field.\newline \indent
In Sec. \ref{Section III}, we include fluctuations of the phase of the order parameter around the saddle point. Our main goal is to study the Berezinskii-Kosterlitz-Thouless (BKT) transition temperature, for which these fluctuations play a crucial role. In Sec. \ref{Section III A}, we start by introducing the phase into our formalism, followed by a derivation of the effective action using the functional-integral adiabatic approximation in Sec. \ref{Section III B}. In Sec. \ref{Section III C}, the resulting effective action is then expanded up to quadratic order in the phase, leading to a phase-fluctuation part of the action. We conclude our calculation by deriving an analytic expression for the fluctuation thermodynamic potential in Sec. \ref{Section III D}. Furthermore, we discuss the effect of spin-orbit coupling on the sound velocities (Sec. \ref{Section III E}) and on the vortex-antivortex structure of the system (Sec. \ref{Section III F}). We then continue to Sec. \ref{Section III G} in which we study the influence of spin-orbit coupling on the BKT critical temperature. Finally, in Sec. \ref{Section III H}, we relate the zero-temperature results to the BKT critical temperature, by discussing the evolution of the quantum phase transition lines at non-zero temperature. In Sec. \ref{IV}, we draw conclusions.
\section{Functional-integral description at the saddle-point level}\label{Section II}
In this section, we set up the functional-integral formalism at the saddle-point level and we discuss the ground states of the system, as well as the zero-temperature phase diagram.
\subsection{Setting up the formalism}\label{Section II A}
In this work, we use a functional-integral approach to calculate thermodynamic properties of the system. More specifically, we write the partition function as a sum over Grassmann fields $\bar{\psi}$ and $\psi$, weighted by the exponential of the action functional $S$
\begin{equation}\label{Partition function fermions}
\mathcal{Z}=\int \mathcal{D}\bar{\psi}_{\textbf{r},\tau,s} \mathcal{D}\psi_{\textbf{r},\tau,s} \exp\left[-S(\bar{\psi}_{\textbf{r},\tau,s},\psi_{\textbf{r},\tau,s})\right].
\end{equation}
Here, $\textbf{r}=(x,y)$ and $\tau$ indicate position and imaginary time, respectively, while $s=\{\uparrow,\downarrow\}$ denotes the spin-state (spin-up and spin-down) of the spin-1/2 fermions.
The action can be related to the Hamiltonian density $\mathcal{H}$ via a Legendre transformation
\begin{align}
&S(\bar{\psi}_{\textbf{r},\tau,s},\psi_{\textbf{r},\tau,s})\nonumber\\&=\int d\tau \int d\textbf{r}\left(\sum_{s} \bar{\psi}_{\textbf{r},\tau,s} \frac{\partial \psi_{\textbf{r},\tau,s}}{\partial \tau} + \mathcal{H}(\bar{\psi}_{\textbf{r},\tau,s},\psi_{\textbf{r},\tau,s})\right).
\end{align}
Our aim is to study a two-dimensional (2D) Fermi gas, where the spin-orbit coupling is a generic combination of Rashba and Dresselhaus terms. The Hamiltonian density of this system can be divided into three parts: $\mathcal{H}=\mathcal{H}_{\rm{0}}+\mathcal{H}_{\rm{S}}+\mathcal{H}_{\rm{I}}$. Note that for the remainder of this paper we use the units $\hbar=2m=k_{\rm{F}}=1$.\newline \indent The first part of $\mathcal{H}$ is
\begin{equation}
\mathcal{H}_{\rm{0}}=\sum_{s,s'}\bar{\psi}_{\textbf{r},\tau,s}\left[\left(-\nabla^2_{\textbf{r}}-\mu_s \right)\delta_{s,s'}-h_z \sigma_{z,ss'} \right]\psi_{\textbf{r},\tau,s'},
\end{equation}
corresponding to the single-particle sector. We work in the grand canonical ensemble, hence the use of the Lagrange multiplier $\mu_s$, which is interpreted as a spin-dependent chemical potential, thus allowing for population imbalance. Furthermore, $h_z$ denotes a Zeeman field perpendicular to the (x-y)-plane. Experimentally, this field corresponds to the intensity $\Omega$ of the Raman transition between different hyperfine states: $h_z=-\Omega/2$. Finally, $\sigma_i$ represents the $i^{th}$ Pauli matrix. \newline \indent 
The second part of $\mathcal{H}$ is
\begin{equation}
\mathcal{H}_{\rm{S}}=-2\sum_{s,s'}\bar{\psi}_{\textbf{r},\tau,s}\left(\alpha \hat{k}_x \sigma_{y,ss'} - \gamma \hat{k}_y \sigma_{x,ss'}\right)\psi_{\textbf{r},\tau,s'},
\end{equation}
corresponding to the spin-orbit terms. Here, we have defined $\alpha=(v_{\rm{R}}+v_{\rm{D}})/2$ and $\gamma=(v_{\rm{R}}-v_{\rm{D}})/2$, with $v_{\rm{R}}$ and $v_{\rm{D}}$ being the Rashba and Dresselhaus coupling strength, respectively. Moreover, $\hat{k}_l=-i(\partial/\partial l)$ is the momentum operator along the $l$-direction.\newline \indent 
The third part $\mathcal{H}_{\rm{I}}$ describes the interaction between fermions. We consider s-wave scattering, thus only taking into account interaction between fermions in different spin states. For a general two-body potential $V(\textbf{r}-\textbf{r}')$, the interaction term can be written as
\begin{equation}
\mathcal{H}_{\rm{I}}=\int d\textbf{r}'\bar{\psi}_{\textbf{r},\tau,\uparrow}\bar{\psi}_{\textbf{r}',\tau,\downarrow}V(\textbf{r}-\textbf{r}')\psi_{\textbf{r}',\tau,\downarrow}\psi_{\textbf{r},\tau,\uparrow}.
\end{equation}
However, in this work, we restrict ourselves to short-range interactions, which can be described by the contact potential $V(\textbf{r}-\textbf{r}')=g\delta(\textbf{r}-\textbf{r}')$. \newline \indent
Having set up the functional-integral formalism, we are ready to discuss the saddle-point approximation.
\subsection{Calculating the saddle-point thermodynamic potential}\label{Section II B}
In this section, our goal is to derive the saddle-point thermodynamic potential from the partition function shown in Eq. (\ref{Partition function fermions}). The difficulty in calculating the latter expression analytically lies in the fourth-order interaction term. A frequently used method to circumvent this problem is to use the Hubbard-Stratonovich transformation
\begin{equation}\label{Hubbard-Stratonovich}
\begin{aligned}
&\exp\left(-g\int d r \bar{\psi}_{r,\uparrow}\bar{\psi}_{r,\downarrow}\psi_{r,\downarrow}\psi_{r,\uparrow}\right)=\int \mathcal{D}\bar{\Delta}_{r}\mathcal{D}\Delta_{r}\\
&\times\exp\left[\int d r \left(\frac{\bar{\Delta}_{r}\Delta_{r}}{g} - \bar{\psi}_{r,\uparrow}\bar{\psi}_{r,\downarrow}\Delta_{r} - \psi_{r,\downarrow}\psi_{r,\uparrow}\bar{\Delta}_{r}\right) \right],
\end{aligned}
\end{equation}
where we denote $r=\{\textbf{r},\tau\}$. This transformation decouples the fourth-order interaction term in Eq. (\ref{Partition function fermions}) into second-order terms, at the cost of inserting an additional functional integral over complex fields $\bar{\Delta}_{\textbf{r},\tau}$ and $\Delta_{\textbf{r},\tau}$. However, these fields have a physical meaning: they are interpreted as the fermion pair fields. In Eq. (\ref{Hubbard-Stratonovich}), we use the Bogoliubov channel in the Hubbard-Stratonovich transformation. It is also possible to use a different channel by using fields which represent the total density (Hartree channel) or the population imbalance density (Fock channel). However, for the description of superfluidity, the Bogoliubov channel is the most natural.\newline \indent
At this point, the fermionic functional integrals in the partition function can be calculated analytically, since the action has been made quadratic in the fermionic fields. It is not possible, however, to calculate the bosonic functional integrals analytically and one has to resort to approximations. Several `levels' of approximation can be considered, starting with the crudest one: the saddle-point approximation. In this case, we assume that the order parameter is constant in time and space or, equivalently, that only its zero momentum component contributes
\begin{align}\label{saddle-point approximation}
\Delta_{\textbf{q},\omega_n}=\sqrt{\beta L^2}\delta(\textbf{q})\delta_{\omega_n,0}|\Delta|.
\end{align}
Hence, we assume that fermions pair at opposite momenta, ignoring the possibility of non-uniform superfluid phases \cite{FFLO Jeroen 1, FFLO Jeroen 2, FFLO Hu, FFLO Zhang}. In Eq. (\ref{saddle-point approximation}), the factor $\sqrt{\beta L^2}$ is added to give $|\Delta|$ dimensions of energy, where $\beta$ denotes inverse temperature and $L^2$ is the area of the 2D system. \newline \indent Fourier transforming the action and applying the saddle-point approximation, the action can be written as
\begin{align}\label{Action saddle-point approximation}
S&=\frac{1}{2}\sum_{\textbf{k},\omega_{n}}\bar{\eta}_{\textbf{k},\omega_{n}}[-i\omega_n\mathbb{I}+\mathcal{H}(\textbf{k})]\eta_{\textbf{k},\omega_{n}}\nonumber\\&+\frac{\beta}{2}\sum_{\textbf{k},\omega_n}\sum_s\left(i\omega_n+\textbf{k}^2-\mu_s \right)-\beta L^2\frac{|\Delta|^{2}}{g},
\end{align}
where $\omega_n=(2n+1)\pi/\beta$ is the fermionic Matsubara frequency, $\textbf{k}$ is the fermionic wave vector and $\mu_s$ is the chemical potential in spin state $s$. However, for the remainder of this paper, we choose $\mu_s=\mu$ and treat only a system with initial identical populations.
In Eq. (\ref{Action saddle-point approximation}) the fermion fields are ordered using the spinor notation
\begin{align}\label{Nambu spinor 4x4}
\bar{\eta}_{\textbf{k},\omega_{n}}=
\begin{pmatrix}
\bar{\psi}_{\textbf{k},\omega_{n},\uparrow} & \bar{\psi}_{\textbf{k},\omega_{n},\downarrow} & \psi_{\textbf{-k},-\omega_{n},\uparrow} & \psi_{\textbf{-k},-\omega_{n},\downarrow}
\end{pmatrix}.
\end{align}
In this basis, the division into a quasiparticle-quasihole part and a spin-up/spin-down part is visible.
The Hamiltonian density appearing in Eq. (\ref{Action saddle-point approximation}) is
\begin{align}\label{Inverse Green's function after saddle-point approximation}
\mathcal{H}(\textbf{k})=
\begin{pmatrix}
\xi_{\textbf{k}}-h_z & -h^*_{\bot}(\textbf{k}) & 0 & |\Delta| \\
-h_{\bot}(\textbf{k})& \xi_{\textbf{k}}+h_z & -|\Delta| & 0 \\
0 & -|\Delta|& -\xi_{\textbf{k}}+h_z & -h_{\bot}(\textbf{k}) \\
|\Delta| & 0 & -h^*_{\bot}(\textbf{k}) &-\xi_{\textbf{k}}-h_z \\
\end{pmatrix}.
\end{align}
The emergence of the second term in Eq. (\ref{Action saddle-point approximation}) stems from the fact that operators have to be Weyl-ordered before they can be mapped onto Grassmann variables. This leads to additional terms due to the anti-commuting nature of the fermion operators. \newline\indent In the matrix shown in Eq. (\ref{Inverse Green's function after saddle-point approximation}), $\xi_\textbf{k}=\textbf{k}^2-\mu$ is the single-particle energy relative to the chemical potential, and $h_\bot({\textbf{k}})=h_x(\textbf{k})+i h_y(\textbf{k})$ is the spin-orbit field with components $h_x(\textbf{k})=-2\gamma k_y$ and $h_y(\textbf{k})=2\alpha k_x$. It is noteworthy to mention that the Hamiltonian density can be written in terms of the Pauli matrices as
\begin{align}
\mathcal{H}(\textbf{k})&=\tau_z \otimes \left(\xi_\textbf{k}\sigma_0 - h_z \sigma_z \right)-|\Delta| \tau_y \otimes \sigma_y \nonumber\\ &+2\gamma k_y \tau_0 \otimes \sigma_x  - 2\alpha k_x \tau_z \otimes \sigma_y,
\end{align}
where the Pauli matrices $\sigma_i$ and $\tau_i$ are associated with the spin-part and the particle-hole part, respectively. Using this notation, it can be shown that quasiparticle-quasihole symmetry is preserved because
\begin{align}
\tau_x \otimes \sigma_0 \mathcal{H}(\textbf{k}) \tau_x \otimes \sigma_0 = -\mathcal{H}^*(-\textbf{k}).
\end{align}
\indent Now, the fermionic functional integral can be calculated, leading to the partition function
\begin{align}\label{Partition function before Mastubara summation}
\mathcal{Z}=& \exp\Bigg(\frac{1}{2}\sum_{\textbf{k},\omega_{n}}{\rm Tr}\left[\ln\left(-\beta\mathbb{G}^{-1}_{\textbf{k},\omega_{n}}\right)\right]\nonumber\\
&-\beta\sum_{\textbf{k},\omega_n}\left(i\omega_n+\textbf{k}^2-\mu \right)+\beta L^2\frac{|\Delta|^{2}}{g} \Bigg).
\end{align}
Finally, performing the Matsubara summation, and replacing the interaction strength $g$ in terms of the two-body binding energy $E_{\rm{B}}$ via the relation
\begin{align}\label{g as a function of E_B}
\frac{1}{g}=-\int\frac{d\textbf{k}}{(2\pi)^2}\frac{1}{2k^2+E_{\rm{B}}},
\end{align}
results in the saddle-point thermodynamic potential
\begin{widetext}
\begin{align}\label{saddle-point free energy}
\Omega_{\rm{sp}}=-\int \frac{d\textbf{k}}{(2\pi)^2}\hspace{1mm}\bm{\left[}\frac{1}{2\beta}\bm{\left(}\ln\left\{2+2\cosh\left[\beta\epsilon_{\rm{p}}^{(+)}(\textbf{k})\right]\right\}+\ln\left\{2+2\cosh\left[\beta\epsilon_{\rm{p}}^{(-)}(\textbf{k})\right]\right\}\bm{\right)}-\xi_\textbf{k}-\frac{|\Delta|^2}{2k^2+E_{\rm{B}}} \bm{\right]}.
\end{align}
\end{widetext}
In Eq. (\ref{saddle-point free energy}), the momentum-dependent functions
\begin{align}\label{quasiparticle energies}
\epsilon_{\rm{p}}^{(\pm)}(\textbf{k})=\sqrt{E_\textbf{k}^2+h_z^2+|h_\bot(\textbf{k})|^2\pm 2\sqrt{E_\textbf{k}^2h_z^2+\xi_\textbf{k}^2 |h_\bot(\textbf{k})|^2}},
\end{align}
represent the quasiparticle energies. Note that by using Eq. (\ref{g as a function of E_B}), we deliberately choose $E_B$ to represent the two-body binding energy in the absence of spin-orbit coupling. In this way, $E_B$ can be treated as an independent system parameter, which makes it easier to identify the direct effects of spin-orbit coupling. For a detailed study of the effect of spin-orbit coupling on the bound state energies resulting from interaction between spin-1/2 fermions, we refer to \cite{Doga Carlos arXiv}. \newline \indent
Having discussed the saddle-point thermodynamic potential, we investigate next the generalized helicity basis.
\subsection{The generalized helicity basis}\label{Section II C}
To gain insight into the effects induced by the presence of spin-orbit coupling, it is instructive to transform the system to the generalized helicity basis, using a momentum dependent SU(2) transformation. In this basis, the Hamiltonian density \textit{in the non-interacting limit} $[\mathcal{H}(\textbf{k})]_{|\Delta|=0}$ becomes diagonal. The transformation matrix is given by
\begin{equation}
U=
\begin{pmatrix}
u_\textbf{k} & v_\textbf{k} & 0 & 0\\
-v^*_\textbf{k} & u_\textbf{k} & 0 & 0\\
0 & 0 & u_{-\textbf{k}} & v^*_{-\textbf{k}}\\
0 & 0 & -v_{-\textbf{k}} & u_{-\textbf{k}} 
\end{pmatrix},
\end{equation}
with the components of the eigenvectors being equal to
\begin{equation}\label{u v (2)}
\begin{array}{l}
u_\textbf{k}=\sqrt{\dfrac{1}{2}\left(1+\dfrac{h_z}{\sqrt{h_z^2+|h_\bot(\textbf{k})|^2}} \right)},\\
v_\textbf{k}=-e^{-i\varphi_\textbf{k}}\sqrt{\dfrac{1}{2}\left(1-\dfrac{h_z}{\sqrt{h_z^2+|h_\bot(\textbf{k})|^2}} \right)}.
\end{array}
\end{equation}
Here, the phase $\varphi_\textbf{k}$ is defined by $h_\bot(\textbf{k})=|h_\bot(\textbf{k})|e^{i\varphi_\textbf{k}}$. Applying the generalized helicity basis transformation to the full Hamiltonian density induces new components of the order parameter
\begin{align}\label{Hamiltonian in helicity basis}
U^\dagger\mathcal{H}(\textbf{k})U=
\begin{pmatrix}
\xi_\Uparrow(\textbf{k}) & 0 & \Delta_{\Uparrow\Uparrow}(\textbf{k}) & \Delta_{\Uparrow\Downarrow}(\textbf{k})\\
0 & \xi_\Downarrow(\textbf{k}) & \Delta_{\Downarrow\Uparrow}(\textbf{k}) & \Delta_{\Downarrow\Downarrow}(\textbf{k})\\
\Delta^*_{\Uparrow\Uparrow}(\textbf{k}) & \Delta^*_{\Downarrow\Uparrow}(\textbf{k}) & -\xi_\Uparrow(\textbf{k}) & 0 \\
\Delta^*_{\Uparrow\Downarrow}(\textbf{k}) & \Delta_{\Downarrow\Downarrow}^*(\textbf{k}) & 0 & -\xi_\Downarrow(\textbf{k})
\end{pmatrix}.
\end{align}
\indent On the diagonal, the energies of the two generalized helicity bands are given by
\begin{equation}
\begin{array}{l}
\xi_\Uparrow(\textbf{k})=\xi_\textbf{k}-\sqrt{h_z^2+|h_\bot(\textbf{k})|^2},\\
\xi_\Downarrow(\textbf{k})=\xi_\textbf{k}+\sqrt{h_z^2+|h_\bot(\textbf{k})|^2}.
\end{array}
\end{equation}
Furthermore, the order parameter in the generalized helicity basis is now a tensor with components
\begin{equation}
\begin{array}{l}
\Delta_{\Uparrow\Uparrow}(\textbf{k})=-\widetilde{\Delta}_{\rm{T}}(\textbf{k}),\\
\Delta_{\Uparrow\Downarrow}(\textbf{k})=\widetilde{\Delta}_{\rm{S}}(\textbf{k}),\\
\Delta_{\Downarrow\Uparrow}(\textbf{k})=-\Delta_{\Uparrow\Downarrow}(\textbf{k}),\\
\Delta_{\Downarrow\Downarrow}(\textbf{k})=\Delta^*_{\Uparrow\Uparrow}(\textbf{k}).
\end{array}
\end{equation}
Here, we identify, respectively, the singlet and the triplet component of the order parameter
\begin{equation}
\begin{array}{l}
\widetilde{\Delta}_{\rm{S}}(\textbf{k})=\dfrac{h_z}{\sqrt{h_z^2+|h_\bot(\textbf{k})|^2}}\hspace{1mm}|\Delta|,\\
\widetilde{\Delta}_{\rm{T}}(\textbf{k})=\dfrac{h^*_\bot(\textbf{k})}{\sqrt{h_z^2+|h_\bot(\textbf{k})|^2}}\hspace{1mm}|\Delta|.
\end{array}
\end{equation}
However, these two components are not independent, as they satisfy
\begin{equation}
|\widetilde{\Delta}_{\rm{S}}(\textbf{k})|^2+|\widetilde{\Delta}_{\rm{T}}(\textbf{k})|^2=|\Delta|^2.
\end{equation}
\indent At this point, it is important to emphasize that our system only has one order parameter, $\Delta$, which is a complex scalar, because there is only s-wave interaction in the original spin-basis. It is only in the generalized helicity basis that the order parameter can be decomposed into a singlet and triplet component and that a spinorial structure arises \cite{3D ERD SOC 2}. The importance of the spinorial structure was also discussed recently in the context of a repulsive Fermi gas \cite{Paper Toigo}. For non-zero spin-orbit coupling, this triplet component cannot be fully suppressed by a Zeeman field, as it involves pairing between particles of equal generalized helicity. Hence, irrespective of the magnitude of the Zeeman field, the order parameter will always contain a triplet component. \newline \indent
We thus see that transforming to the generalized helicity basis has effectively changed the local isotropic s-wave interaction into a non-local anisotropic interaction. The triplet component of the order parameter is not inherently present in the system, instead it is induced by the spin-orbit coupling. For a related discussion on the singlet and triplet components of the condensate fraction we refer to \cite{Salasnich singlet triplet}. \newline \indent Now that we have presented a detailed discussion of the generalized helicity basis, we continue by studying the various uniform superfluid phases of the system. 
\subsection{Topologically distinct superfluid phases}\label{Section II D}
\begin{figure}[tb]
\centering
\includegraphics[keepaspectratio=true,width=86mm]{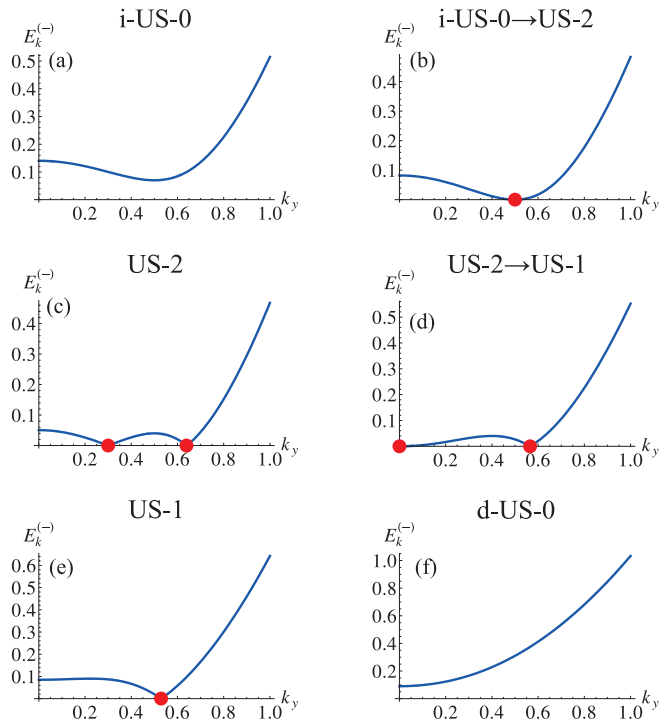}
\caption{Overview of the topologically distinct uniform superfluid (US) phases, categorized by the nodal structure of the lowest quasiparticle energy branch. In Figs. (a),(b),(c) and (d), the values of $\mu$ and $|\Delta|$ are held fixed while $h_z$ is increased. In Figs. (e) and (f), the values of $|\Delta|$ and $h_z$ are held fixed while $\mu$ is decreased. Note that these phases only occur in the ERD case.}
\label{Figure: US-phases}
\end{figure}

The Hamiltonian density of the system has four eigenvalues: two quasiparticle energies $\epsilon_{\rm{p}}^{(\pm)}(\textbf{k})$, shown in Eq. (\ref{quasiparticle energies}), and two quasihole energies $\epsilon_{\rm{h}}^{(\pm)}(\textbf{k})=-\epsilon_{\rm{p}}^{(\pm)}(\textbf{k})$. The structure of these excitation spectra can be used to distinguish different uniform superfluid (US) phases. Let us look more closely at the quasiparticle branches. The (+)-branch is always gapped, whereas the (-)-branch can have nodes depending on the system parameters. In the language of the generalized helicity basis, we can write the second quasiparticle energy as
\begin{align}\label{(-) branch quasiparticle energy}
\epsilon_{\rm{p}}^{(-)}(\textbf{k})=\sqrt{[E_{\rm{S}}(\textbf{k})-|\textbf{h}_{\rm{eff}}(\textbf{k})|]^2+|\widetilde{\Delta}_{\rm{T}}(\textbf{k})|^2}.
\end{align}
Here, we have introduced the energy associated with the singlet channel $E_{\rm{S}}(\textbf{k})=\sqrt{\xi_\textbf{k}^2+|\widetilde{\Delta}_{\rm{S}}(\textbf{k})|^2}$, as well as an effective magnetic field $\textbf{h}_{\rm{eff}}(\textbf{k})=(h_\bot(\textbf{k}),h_z)$ which is a combination of the spin-orbit and the Zeeman fields. This effective field can also be written as half the energy difference of the generalized helicity bands $|\textbf{h}_{\rm{eff}}|=[\xi_\Downarrow(\textbf{k})-\xi_\Uparrow(\textbf{k})]/2$, while the single-particle energy is equal to the average energy of the helicity bands $\xi_\textbf{k}=[\xi_\Uparrow(\textbf{k})+\xi_\Downarrow(\textbf{k})]/2$. \newline \indent The lowest quasiparticle energy branch $\epsilon_{\rm{p}}^{(-)}(\textbf{k})$ has nodes whenever the following two conditions are satisfied simultaneously: 1) the effective magnetic field $\textbf{h}_{\rm{eff}}(\textbf{k})$ and the singlet energy $E_{\rm{S}}(\textbf{k})$ are equal in magnitude, and 2) the triplet component $\widetilde{\Delta}_{\rm{T}}(\textbf{k})$ of the order parameter is zero. In the ERD case, $\widetilde{\Delta}_{\rm{T}}(\textbf{k})=0$ leads to $k_x=0$, which together with $E_{\rm{S}}(\textbf{k})=|h_{\rm{eff}}(\textbf{k})|$ gives the relation $(k_y^2-\mu)^2+|\Delta|^2=h_z^2$, yielding the possibility of having nodes at non-zero momentum. By contrast, in the RO case, or any other combination of Rashba and Dresselhaus terms, no nodes are present at non-zero momentum. \newline \indent The different possible phases in the ERD case are shown in Fig. \ref{Figure: US-phases}. Let us describe these phases in more detail. When the Zeeman field is smaller than the order parameter ($h_z<|\Delta|$), two phases can be discerned: 1) If the chemical potential $\mu>0$, the system acquires an indirect gap at non-zero $|\Delta|$, and 2) if $\mu<0$ a direct gap at $k_y=0$ occurs. These phases are labeled i-US-0 and d-US-0, respectively. When the Zeeman field becomes larger than the order parameter ($h_z>|\Delta|$), the quasiparticle spectrum acquires a nodal structure, depending on the value of the chemical potential. If $\mu>\sqrt{h_z^2-|\Delta|^2}$, the spectrum has two pairs of nodes (US-2 phase). When $|\mu|<\sqrt{h_z^2-|\Delta|^2}$, one pair of nodes is removed from the spectrum at $k=0$ and only one pair remains (US-1 phase). Finally, when $\mu<-\sqrt{h_z^2-|\Delta|^2}$ the final pair of nodes also vanishes at $k=0$ and the system becomes directly gapped (d-US-0 phase). To summarize, the different topological phases can be classified as follows
\begin{equation}\label{US-phases overview}
\begin{array}{l}
h_z<|\Delta|\rightarrow
\begin{dcases}
\mu>0\quad\text{i-US-0}\\
\mu<0\quad\text{d-US-0}
\end{dcases}\\
h_z>|\Delta|\rightarrow
\begin{dcases}
\mu>\sqrt{h_z^2-|\Delta|^2}\quad\text{US-2}\\
|\mu|<\sqrt{h_z^2-|\Delta|^2}\quad\text{US-1}\\
\mu<-\sqrt{h_z^2-|\Delta|^2}\quad\text{d-US-0}
\end{dcases}
\end{array}.
\end{equation}
\newline \indent
There are several effects induced by the nodal structure of the quasi-particles energies at low temperatures ($T << T_{\rm{BKT}}$). First and foremost there is a dramatic change in the momentum distribution of the system. For example, in the i-US-0 phase the momentum distribution is a smooth function, whereas in the US-2 phase discontinuities develop. Furthermore, both the isothermal compressibility and the spin susceptibility are non-analytic at the phase boundaries between the different uniform superfluid phases. This provides clear thermodynamic signatures of the quantum phase transitions. For more details we refer to \cite{Li Carlos arXiv}. Furthermore, in this paper, the emergence of nodes in the order parameter, when viewed in the generalized helicity basis, leads to anisotropies in the superfluid density tensor and sound velocities, as described in sections III C and III E.\newline \indent
Having identified the topological nature of the uniform superfluid phases, we continue by discussing the ground state phase diagram.
\subsection{Zero-temperature phase diagram}\label{Section II E}
\begin{figure}[tb]
\centering
\includegraphics[keepaspectratio=true,width=60mm]{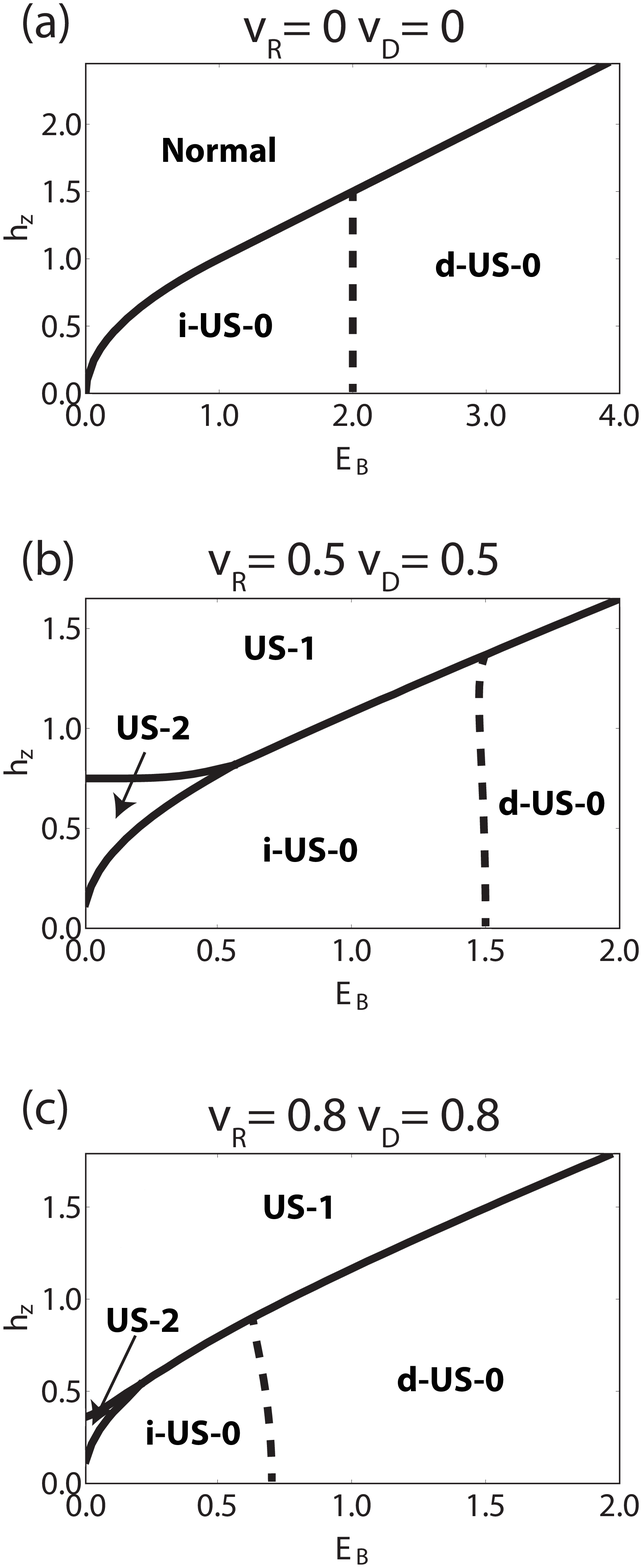}
\caption{Zero-temperature phase diagram, as a function of the two-body binding energy $E_{\rm{B}}$ and a Zeeman field $h_z$, for different values of the spin-orbit coupling strength. In these figures we consider only equal-Rashba-Dresselhaus spin-orbit coupling ($v=v_{\rm{R}}=v_{\rm{D}}$): (a) $v=0$, (b) $v/\tilde{v}_{\rm{F}}=0.5$ and (c) $v/\tilde{v}_{\rm{F}}=0.8$ ($\tilde{v}_{\rm{F}}=v_{\rm{F}}/2$ with $v_{\rm{F}}$ the Fermi velocity).}
\label{Figure: Ebhz-phase diagram}
\end{figure}

\begin{figure}[tb]
\centering
\includegraphics[keepaspectratio=true,width=86mm]{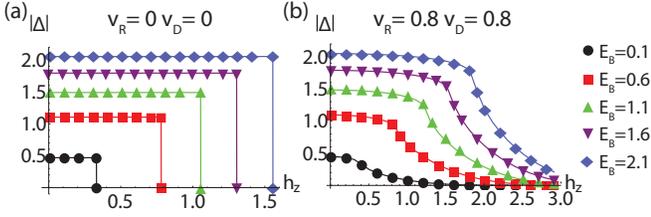}
\caption{Order parameter as a function of the Zeeman field $h_z$. In the case without spin-orbit coupling, there is a first order phase transition at a given critical Zeeman field. In contrast, when (ERD) spin-orbit coupling is present, $|\Delta|$ only goes to zero in the limit $h_z \rightarrow \infty$.}
\label{Figure: Delta_ifv_hz}
\end{figure}

To find out which of the aforementioned superfluid phases occur, we investigate the zero-temperature phase diagram as a function of the two-body binding energy $E_{\rm{B}}$ and Zeeman field $h_z$. More specifically, for a given $(E_{\rm{B}},h_z)$-point, we minimize the saddle-point free energy $F_{\rm{sp}}=\Omega_{\rm{sp}}+\mu n$ with respect to the order parameter $|\Delta|$, while simultaneously solving the number equation $\partial \Omega_{\rm{sp}}/\partial \mu = -n$ in order to determine the chemical potential $\mu$. The resulting values of $|\Delta|$ and $\mu$ then determine which phase the system reaches, according to Eq. (\ref{US-phases overview}). In Fig. \ref{Figure: Ebhz-phase diagram}, the resulting $(E_{\rm{B}},h_z)$-phase diagram is shown for several values of the spin-orbit coupling strength.
\newline \indent Figure \ref{Figure: Ebhz-phase diagram}(a) shows the case without spin-orbit coupling. In this case, only the standard gapped superfluid phase (US-0) occurs, with a crossover between an indirect gap (i-US-0) at low binding energy and a direct gap (d-US-0) at large binding energy. At each value of the binding energy a critical Zeeman field exists at which a first order transition occurs from the US-0 phase to the normal phase. The driving mechanism behind this transition is the energy splitting between spin-up and spin-down fermions caused by the Zeeman field, which suppresses spin-singlet pairing at opposite momenta. The stronger the two-body binding energy between fermions, the larger the Zeeman field must be to break up the fermion pairs. This first order phase transition is also visible in Fig. \ref{Figure: Delta_ifv_hz}(a), where the order parameter $|\Delta|$ is shown as a function of $h_z$, for several values of the binding energy. This figure reveals that the order parameter jumps discontinuously to zero at a critical Zeeman field which depends on the value of $E_{\rm{B}}$.\newline \indent
Figures \ref{Figure: Ebhz-phase diagram}(b) and (c) show the case of ERD spin-orbit coupling with $v_{\rm{R}}/\tilde{v}_{\rm{F}}=v_{\rm{D}}/\tilde{v}_{\rm{F}}=0.5$ and $v_{\rm{R}}/\tilde{v}_{\rm{F}}=v_{\rm{D}}/\tilde{v}_{\rm{F}}=0.8$, respectively (with $\tilde{v}_{\rm{F}}=v_{\rm{F}}/2$ and $v_{\rm{F}}$ the Fermi velocity). These figures demonstrate the existence of the topological uniform superfluid phases US-1 and US-2 at zero temperature.  An important difference between the case without spin-orbit coupling and the ERD case is that in the former the system always transitions into the normal phase at high Zeeman field, whereas in the latter the system transitions into the US-1 phase. The origin of this difference lies in the triplet component of the order parameter that is induced by the presence of spin-orbit coupling, as demonstrated in section \ref{Section II C}. This triplet component cannot be suppressed by a Zeeman field, as it involves pairing between particles of equal generalized helicity. Hence, irrespective of the magnitude of the Zeeman field, the order parameter will always contain a triplet component and as a result will only become zero in the limit $h_z \rightarrow \infty$. This behavior is shown explicitly in Fig. \ref{Figure: Delta_ifv_hz}(b): at low values of the Zeeman field, the order parameter is approximately constant, while at high values it becomes suppressed but stays non-zero.  Figures \ref{Figure: Ebhz-phase diagram} and \ref{Figure: Delta_ifv_hz} both coincide perfectly with the results of \cite{Li Carlos arXiv}. \newline \indent  Figures \ref{Figure: Ebhz-phase diagram}(b) and (c) both show a triple point, in which the i-US-0 phase, the US-2 phase and the US-1 phase meet. For low binding energy, the system undergoes two phase transitions with increasing Zeeman field: i-US-0 $\rightarrow$ US-2 $\rightarrow$ US-1.
At higher binding energy, the system undergoes only one phase transition: US-0 $\rightarrow$ US-1. With increasing binding energy, the region of the US-2 phase shrinks because the lower bound increases. This occurs because the gap in the quasiparticle spectrum of the i-US-0 phase increases with increasing binding energy, and thus a higher value of the Zeeman field is needed in order to bridge this gap. Note also that this gap only disappears at $k_x=0$ in momentum space. For all other values of $\textbf{k}$ the gap is topologically protected by the presence of spin-orbit coupling. Here, we conclude our discussion of the saddle-point case and move on to include phase fluctuations around the saddle point. 
\section{Phase fluctuations around the saddle point}\label{Section III}
In this section, we discuss the effects of phase fluctuations and its impact on the finite temperature phase diagram, sound velocities and vortex-antivortex structure.
\subsection{Introducing the phase}\label{Section III A}
The scope of this work is to study the Berezinskii-Kosterlitz-Thouless (BKT) transition, in which phase fluctuations of the order parameter play a fundamental role. To introduce the phase, the complex field of the order parameter can be re-written as
\begin{equation}\label{inserting the phase}
\Delta_{\textbf{r},\tau}=|\Delta_{\textbf{r},\tau}|e^{i\theta_{\textbf{r},\tau}}.
\end{equation}
Furthermore, we use the gauge transformation $\psi_{\textbf{r},\tau}\rightarrow \psi_{\textbf{r},\tau}e^{i\theta_{\textbf{r},\tau}/2}$ to make explicit the dependence of the action on the phase.
\begin{widetext}
Inserting Eq. (\ref{inserting the phase}) into the partition function of Eq. (\ref{Partition function fermions}) after applying the Hubbard-Stratonovich transformation defined in Eq. (\ref{Hubbard-Stratonovich}) yields
\begin{equation}\label{Partition function fermions bosons amplitude phase}
\begin{aligned}
\mathcal{Z}&=\int\mathcal{D}\bar{\psi}_{\textbf{r},\tau,s}\mathcal{D}\psi_{\textbf{r},\tau,s}\int\mathcal{D}|\Delta_{\textbf{r},\tau}|\mathcal{D}\theta_{\textbf{r},\tau}\exp\left(-\int d\tau \int d\textbf{r}
\left[I_{\rm{0}}(\textbf{r},\tau)+I_{\rm{S}}(\textbf{r},\tau)+I_{\rm{I}}(\textbf{r},\tau) \right] \right),
\end{aligned}
\end{equation}
where the different parts of the action (single-particle, spin-orbit coupling and interaction) are given by
\begin{equation}
\begin{array}{l}
I_{\rm{0}}(\textbf{r},\tau)=\sum_s\bar{\psi}_{\textbf{r},\tau,s}\left( \dfrac{\partial}{\partial\tau}-\nabla^{2}_{\textbf{r}}-\mu+\dfrac{i}{2}\dfrac{\partial\theta_{\textbf{r},\tau}}{\partial\tau}-i[\nabla_{\textbf{r}}(\theta_{\textbf{r},\tau})]\cdot\nabla_{\textbf{r}}-\dfrac{i}{2}\nabla^{2}_{\textbf{r}}(\theta_{\textbf{r},\tau})+\dfrac{1}{4}[\nabla_{\textbf{r}}(\theta_{\textbf{r},\tau})]^{2} \right)\psi_{\textbf{r},\tau,s},\\
I_{\rm{S}}(\textbf{r},\tau)=2\sum_{s}\bar{\psi}_{\textbf{r},\tau,s}\left[s\alpha\left(\dfrac{\partial}{\partial x}+\dfrac{i}{2}\dfrac{\partial\theta_{\textbf{r},\tau}}{\partial x} \right)-i\gamma\left(\dfrac{\partial}{\partial y}+\dfrac{i}{2}\dfrac{\partial\theta_{\textbf{r},\tau}}{\partial y}\right)\right]\psi_{\textbf{r},\tau,-s},\\
I_{\rm{I}}(\textbf{r},\tau)=\bar{\psi}_{\textbf{r},\tau,\uparrow} \bar{\psi}_{\textbf{r},\tau,\downarrow}\Delta_{\textbf{r},\tau} + \psi_{\textbf{r},\tau,\downarrow}\psi_{\textbf{r},\tau,\uparrow} \bar{\Delta}_{\textbf{r},\tau} -\dfrac{\bar{\Delta}_{\textbf{r},\tau}\Delta_{\textbf{r},\tau}}{g}.
\end{array}
\end{equation}
\end{widetext}
Here, we used the symbol $s$ ambiguously: $s=\{\uparrow,\downarrow\}$ when used as an index and $s=\pm 1$ when used as a number.\newline
Since phase fluctuations provide the dominant contribution to the physics in 2D, we ignore from this point on the contribution of amplitude fluctuations by assuming that the amplitude of the order parameter is constant: $|\Delta_{\textbf{r},\tau}|=|\Delta|$. This still leaves three functional integrals to be calculated for the partition function described in Eq. (\ref{Partition function fermions bosons amplitude phase}).
\subsection{The adiabatic approximation}\label{Section III B}
 In principle, the fermionic functional integral in Eq. (\ref{Partition function fermions bosons amplitude phase}) can be calculated exactly because the action is quadratic in the fermionic fields. However, we need first to transform the action to reciprocal space in order to eliminate the space-time derivatives. At this point in the calculation, the \textit{functional-integral adiabatic approximation} is used \cite{phase fluctuations 1,phase fluctuations 2,FFLO 3D fluctations Jeroen Jacques}. We assume that the phase field $\theta_{\textbf{r},\tau}$ varies slowly in space and time compared to a similar variation of the fermionic fields $\bar{\psi}_{\textbf{r},\tau,s}$ and $\psi_{\textbf{r},\tau,s}$. As a result, for a given configuration of the phase field, the configuration of fermionic fields can be coarse-grained by averaging over the `fast' degrees of freedom. \newline \indent Given this approximation, we can Fourier transform the partition function in Eq. (\ref{Partition function fermions bosons amplitude phase}) and calculate the fermionic functional integrals analytically. This procedure leads to
\begin{equation}\label{partition function path integral over phase}
\mathcal{Z}=\int \mathcal{D}\theta_{\textbf{r},\tau} \exp\left[-S_{\rm{eff}}(\theta_{\textbf{r},\tau}) \right],
\end{equation}
where the phase-only effective action is given by
\begin{widetext}
\begin{equation}\label{Action after functional integration over Delta and psi}
\begin{aligned}
S_{\rm{eff}}=-\frac{1}{2} {\rm Tr} \left\{\ln\left[\beta\begin{pmatrix}
\mathbb{M}_+&&\mathbb{D}_+\\
\mathbb{D}_-&&\mathbb{M}_-^*
\end{pmatrix}\right]\right\}-\frac{\beta L^2 |\Delta|^2}{g}
+\beta\sum_{\textbf{k},\omega_n}(-i\omega_n+\textbf{k}^2-\mu)+\frac{1}{8 L^2}\int d\tau \int d\textbf{r}\sum_{\textbf{k},\omega_n}[\nabla_\textbf{r}(\theta_{\textbf{r},\tau})]^2.
\end{aligned}
\end{equation}
\end{widetext}
In this expression, $\beta$ is the inverse temperature, $L^2$ is the area of the 2D system, $\textbf{k}$ denotes the fermionic wave vector and $\omega_n=(2n+1)\pi/\beta$ is the fermionic Matsubara frequency. The matrix in Eq. (\ref{Action after functional integration over Delta and psi}) has dimensions $4\times 4$ and can be written in terms of the $2\times 2$ matrices
\begin{equation}\label{submatrix A}
\mathbb{M}_\pm=
\begin{pmatrix}
\mp i\omega_n\pm\xi_\textbf{k}^\theta\mp h_z-\zeta_\textbf{k}^\theta&&-h^*_{\bot}(\textbf{k})\mp h_{\bot}^{\theta}\\
-h_{\bot}(\textbf{k})\mp h_{\bot}^{*\theta}&&\mp i\omega_n\pm\xi_\textbf{k}^\theta\pm h_z-\zeta_\textbf{k}^\theta
\end{pmatrix}
\end{equation}
and $\mathbb{D}_{\pm}=\pm i\sigma_y|\Delta|$. \newline \indent The kinetic terms in Eq. (\ref{submatrix A}) have been divided into phase-independent and phase-dependent contributions, where we defined $\xi_\textbf{k}^\theta=\xi_\textbf{k}+\xi^\theta$. The phase-independent terms are $\xi_\textbf{k}=\textbf{k}^2-\mu$  and $h_z$. The phase-dependent terms are $\xi^\theta=\frac{i}{2}\frac{\partial\theta_{\textbf{r},\tau}}{\partial\tau}+\frac{1}{4}[\nabla_\textbf{r}(\theta_{\textbf{r},\tau})]^2$ and $\zeta_\textbf{k}^\theta=-\nabla_\textbf{r}(\theta_{\textbf{r},\tau})\cdot\textbf{k}$. The spin-flip terms also contain a phase-independent contribution corresponding to the spin-orbit coupling field $h_{\bot}(\textbf{k})=-2\gamma\hspace{0.5mm}k_y+2i\alpha\hspace{0.5mm}k_x$ and a phase-dependent contribution $h_{\bot}^\theta=-\gamma\frac{\partial\theta_{\textbf{r},\tau}}{\partial y}-i\alpha\frac{\partial\theta_{\textbf{r},\tau}}{\partial x}$. \newline
In Eq. (\ref{Action after functional integration over Delta and psi}), the two final terms emerge by taking into account the anti-commuting nature of fermionic operators. This is done by Weyl-ordering these operators in second quantized form, before mapping them onto Grassmann variables.
\subsection{Expansion of the effective action up to quadratic order in the phase}\label{Section III C}
The exact calculation of the partition function shown in Eq. (\ref{Partition function fermions bosons amplitude phase}) requires knowledge of the eigenvalues of the matrix described in Eq. (\ref{Action after functional integration over Delta and psi}). These eigenvalues are the solution of a quartic equation, and hence are too cumbersome to be used in any analytic solution for the effective action. Instead, we treat the spatial and temporal derivatives of the phase-field as a small perturbation and expand the effective action up to second order in terms of these derivatives. \newline\indent Let us first introduce a shorter notation $\mathbb{A}_{\textbf{k},\omega_n}(\theta,\partial \theta)$ for the $4 \times 4$ matrix in Eq. (\ref{Action after functional integration over Delta and psi}). By adding and subtracting the phase-independent part of the effective action, we re-write the trace-log of this matrix as ${\rm Tr}\left\{\ln[\mathbb{A}_{\textbf{k},\omega_n}(0,0)+\mathbb{F}_\textbf{k}(\theta,\partial\theta)] \right\}$, where
\begin{equation}
\mathbb{F}_\textbf{k}(\theta,\partial\theta)=
\begin{pmatrix}
\varepsilon^\theta_+(\textbf{k}) && -h_\bot^{\theta} && 0 && 0\\
-\left(h_\bot^{\theta}\right)^*&& \varepsilon^\theta_+(\textbf{k}) && 0 && 0\\
0 && 0 && \varepsilon^\theta_-(\textbf{k}) && \left(h_\bot^{\theta}\right)^*\\
0 && 0 && h_\bot^{\theta} && \varepsilon^\theta_-(\textbf{k})
\end{pmatrix},
\end{equation}
is the fluctuations part, with $\varepsilon^\theta_{\pm}(\textbf{k})=\pm \xi^\theta - \zeta_\textbf{k}^{\theta}$. Using this notation, the trace-log can be written as
\begin{equation}\label{Tracelog}
\begin{aligned}
&{\rm Tr}\{\ln[\mathbb{A}_{\textbf{k},\omega_n}(0,0)+\mathbb{F}_\textbf{k}(\theta,\partial\theta)]\}\\
=&{\rm Tr}\{\ln[\mathbb{A}_{\textbf{k},\omega_n}(0,0)]\}+{\rm Tr}\{\ln[\mathbb{I}+\mathbb{A}^{-1}_{\textbf{k},\omega_n}(0,0) \mathbb{F}_\textbf{k}(\theta,\partial\theta)]\},
\end{aligned}
\end{equation} 
where the first term on the right-hand side is the saddle-point contribution, and the second term is due to phase fluctuations. To preserve the readability of the paper we refer the details of the explicit calculation of $S_{\rm{eff}}$ to appendix \ref{Appendix: expanding the action}. Here, we just show the final expression for the effective action $S_{\rm{eff}}=S_{\rm{sp}}+S_{\rm{fl}}$, where $S_{\rm{sp}}$ is the saddle-point contribution and
\begin{equation}\label{fluctuation action}
S_{\rm{fl}}=\frac{1}{2}\int d\tau \int d\textbf{r}\left(\mathcal{A}\left(\frac{\partial\theta_{\textbf{r},\tau}}{\partial\tau}\right)^2+\sum_\nu\rho_{\nu\nu}\left( \frac{\partial\theta_{\textbf{r},\tau}}{\partial \nu} \right)^2 \right),
\end{equation}
is the fluctuation action with ${\nu=\{x,y\}}$. \newline \indent 
Due to the presence of anisotropic spin-orbit coupling, the superfluid density tensor has unequal components $\rho_{xx} \neq \rho_{yy}$, except in the isotropic Rashba-only case (as well as the Dresselhaus-only case), where $\rho_{xx}=\rho_{yy}$. Using a simple scale transformation, the action in Eq. (\ref{fluctuation action}) can be written in a form that is equivalent to the action in the case without spin-orbit coupling:
\begin{equation}\label{fluctuation action rescaled}
S_{\rm{fl}}=\frac{1}{2}\int d\tau \int d\textbf{r}\left(\mathcal{A}\left(\frac{\partial\theta_{\textbf{r},\tau}}{\partial\tau}\right)^2+\rho_{\rm{s}} \left[\nabla(\theta_{\textbf{r},\tau})\right]^2 \right),
\end{equation}
where the effective superfluid density is now given by $\rho_{\rm{s}}=\sqrt{\rho_{xx}\rho_{yy}}$. The exact expressions for the compressibility $\mathcal{A}$ and the superfluid density components $\rho_{xx},\rho_{yy}$ are given in appendix \ref{Appendix: expanding the action}.
\subsection{The fluctuation thermodynamic potential}\label{Section III D}
Finally, the fluctuation thermodynamic potential can be derived from the action given in Eq. (\ref{fluctuation action rescaled}). When calculating the functional integral defined in Eq. (\ref{partition function path integral over phase}), one has to be careful not to double count the fields, since $\theta_{\textbf{q},\varpi_m}$ is a real field, hence $\theta^*_{\textbf{q},\varpi_m}=\theta_{-\textbf{q},-\varpi_m}$. Here, $\textbf{q}$ is the bosonic wave vector and $\varpi_m=2\pi m/\beta$ is the bosonic Matsubara frequency. A way to circumvent this problem is to write the partition function over half the total $\textbf{q}$-space:
\begin{equation}
\begin{aligned}
\mathcal{Z}_{\rm{fl}}&=\prod_{\substack{\textbf{q},\varpi_m\\q_x>0}}\int d\theta_{\textbf{q},\varpi_m} d\theta^*_{\textbf{q},\varpi_m}\nonumber\\
&\times\exp\left(-\sum_{\substack{\textbf{q},\varpi_m\\q_x>0}}\left(\mathcal{A}\varpi_m^2+\rho_{\rm{s}} q^2\right)\theta_{\textbf{q},\varpi_m}\theta^*_{\textbf{q},\varpi_m} \right).
\end{aligned}
\end{equation}
Now the known result of a Gaussian bosonic functional integral can be applied, which leads to
\begin{equation}
\mathcal{Z}_{\rm{fl}}=\exp \left(-\frac{1}{2} \sum_{\textbf{q},\varpi_m}\ln\left[\beta^2\left(\mathcal{A}\varpi_m^2+\rho_{\rm{s}} q^2\right)\right] \right).
\end{equation}
After performing the bosonic Matsubara summation, and using $\mathcal{Z}_{\rm{fl}}=\exp(-\beta \Omega_{\rm{fl}})$ we arrive at
\begin{equation}\label{superfluid free energy}
\Omega_{\rm{fl}}=\frac{L^2}{2\pi\beta}\int_0^\infty dq\hspace{1mm}q\ln\left(1-e^{-\beta \omega(\textbf{q})} \right).
\end{equation}
In 2D, the integral in this expression can be calculated analytically, yielding
\begin{equation}
\Omega_{\rm{fl}}=-\zeta(3)\frac{L^2}{2\pi\beta^3}\frac{\mathcal{A}}{\rho_{\rm{s}}},
\end{equation}
which corresponds to the additional pressure
\begin{equation}
\delta P=\frac{\zeta(3)}{2\pi}\frac{\mathcal{A}}{\rho_s}T^3,
\end{equation}
created by exciting collective (sound) modes, to be discussed in the next section.
\subsection{Sound velocities}\label{Section III E}
In the action shown in Eq. (\ref{fluctuation action}), the presence of spin-orbit coupling results in an anisotropic superfluid density tensor (except in the isotropic RO case). As a result, the sound velocities of the system become anisotropic. In the case without spin-orbit coupling, the sound velocity is equal to $c=\sqrt{\rho_s/\mathcal{A}}$, where $\mathcal{A}$ is the compressibility and $\rho_{\rm{s}}$ is the superfluid density. In the present case, the sound velocities in the x- and y-direction are given by
\begin{align}
\begin{matrix}
c_x=\sqrt{\dfrac{\rho_{xx}}{\mathcal{A}}}, && c_y=\sqrt{\dfrac{\rho_{yy}}{\mathcal{A}}},
\end{matrix}
\end{align}
respectively.
\begin{figure}[bt]
\centering
\includegraphics[width=86mm]{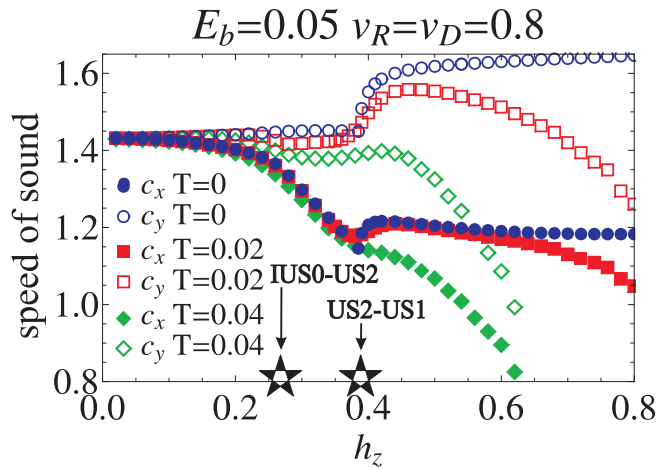}
\caption{Sound velocities as a function of the Zeeman field $h_z$, for a given value of the binding energy $E_{\rm{B}}/E_{\rm{F}}=0.05$ and spin-orbit coupling strength $v_{\rm{R}}/\tilde{v}_{\rm{F}}=v_{\rm{D}}/\tilde{v}_{\rm{F}}=0.8$. The sound velocities are shown at three different temperatures: $T/T_{\rm{F}}$=0, $T/T_{\rm{F}}$=0.02 and $T/T_{\rm{F}}$=0.04. The presence of ERD spin-orbit coupling induces an anisotropy in the sound velocities. Moreover, they are sensitive to the quantum phase transition between the US-2 and US-1 phase \cite{PRL Jeroen Jacques Carlos}. Notice that the vertical axis does not start at zero.}
\label{Figure: sound velocities}
\end{figure}
\indent The anisotropic sound velocities of the system are shown in Fig. \ref{Figure: sound velocities} as a function of the Zeeman field $h_z$, at different temperatures. The binding energy is held fixed at $E_{\rm{B}}/E_{\rm{F}}=0.05$ and we consider ERD spin-orbit coupling with $v_{\rm{R}}/\tilde{v}_{\rm{F}}=v_{\rm{D}}/\tilde{v}_{\rm{F}}=0.8$. This means that we follow a vertical line through the phase diagram in Fig. \ref{Figure: Ebhz-phase diagram}(c). Along this line with increasing $h_z$, the following uniform superfluid phases are encountered: i-US-0,US-2 and US-1. The transition points between these phases are indicated on the abscissa of Fig. \ref{Figure: sound velocities}. In this figure, we see that the sound velocities are sensitive to the presence of the quantum phase transition between the US-2 and US-1 phases: at this transition point, both sound velocities show a distinct cusp. On the other hand, the sound velocities do not show such behavior at the transition point between i-US-0 and US-2. \newline \indent The reason for this difference in sensitivity is that at the US-2-to-US-1 transition, two nodal Dirac quasiparticles with opposite topological charges annihilate at zero momentum, i.e. in the long-wavelength limit. Because sound waves are low-energy and long-wavelength excitations, they tend to be sensitive to this transition at $\textbf{k}=0$. The i-US-0-to-US-2 transition, however, can be understood (when approached from the US-2 side) as the annihilation of two Dirac quasiparticles
with opposite topological charges at \textit{non-zero} momentum. Therefore, the sound velocities are much less sensitive to this quantum phase transition.
\newline\indent As expected, the sharpness of the cusps in the sound velocities tends to soften when temperature is increased. Note also that in the limit of $h_z \rightarrow 0$, both sound velocities converge to the known limit $c_x=c_y=v_{\rm{F}}/\sqrt{2}$ of the case without spin-orbit coupling. This shows that in the ERD case, the effect of spin-orbit coupling can be gauged away in the absence of the Zeeman field $h_z$. Now that the sound wave excitations have been discussed, we proceed by analyzing the topological excitations.

\subsection{Vortex-antivortex structure}\label{Section III F}
\begin{figure}[bt]
\centering
\includegraphics[width=86mm]{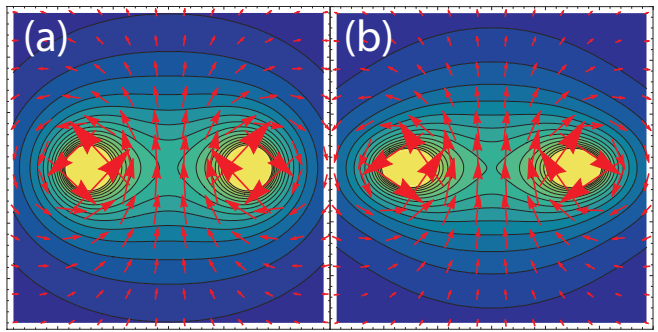}
\caption{Vortex-antivortex structure for a 2D Fermi gas in (a) the Rashba-only case and (b) the equal Rashba-Dresselhaus case. In the presence of ERD spin-orbit coupling, the vortices and antivortices become elliptic. In the isotropic Rashba-only case, the vortex structure remains circular. The parameters used are $h_z/E_{\rm{F}}=0.2$, $E_{\rm{B}}/E_{\rm{F}}=0.01$, $T\approx T_{\rm{BKT}}$, with $v_{\rm{R}}/\tilde{v}_{\rm{F}}=1$ in (a) and $v_{\rm{R}}/\tilde{v}_{\rm{F}}=v_{\rm{D}}/\tilde{v}_{\rm{F}}=1$ in (b). \cite{PRL Jeroen Jacques Carlos}.}
\label{VortexAntivortex.eps}
\end{figure}

The anisotropy of the superfluid density also has repercussions on the vortex and antivortex structure of the system. The vortex solutions are found by extremizing the fluctuation action (\ref{fluctuation action rescaled}) in the static case ($\tau=0$). This leads to Laplace's equation $\nabla^2(\theta_{\textbf{r}})=0$, which has singular, vortex-like solutions that are given by $\theta_{\pm}(x,y)=\pm\arctan(y/x)$. However, since the original fluctuation-action is given in Eq. (\ref{fluctuation action}), we still need to transform the solution in order to reflect the rescaling that was performed in order to transform Eq. (\ref{fluctuation action}) into Eq. (\ref{fluctuation action rescaled}). This leads to the solution
\begin{align}
\theta_\pm(x,y)=\pm\arctan\left(\sqrt{\frac{\rho_{xx}}{\rho_{yy}}}\frac{y}{x} \right),
\end{align}
for a vortex($+$)/antivortex($-$) located at $(x=0,y=0)$. This shows that the vortices(V) and antivortices(A) present in the system become elliptical, rather than circular, in the presence of ERD spin-orbit coupling. For the isotropic Rashba-only case, vortices remain circular. \newline \indent The general solution for a vortex-antivortex (VA) pair is found by taking the sum of two vortex-solutions with opposite winding number, leading to
\begin{equation}\label{VA solution}
\theta_{\rm{VA}}(x,y)=\arctan\left(\frac{2\tilde{a}\tilde{y}}{\tilde{a}^2-\tilde{x}^2-\tilde{y}^2}\right),
\end{equation}
where $\tilde{x}=x/\tilde{\rho}$, $\tilde{y}=y\tilde{\rho}$ and $\tilde{a}=a/\tilde{\rho}$, with $\tilde{\rho}=(\rho_{xx}/\rho_{yy})^{1/4}$. The parameter $a$ indicates the distance between the cores of a vortex and an antivortex located at $(x=-a,y=0)$ and $(x=a,y=0)$, respectively. Plots of the solution given by Eq. (\ref{VA solution}) are shown in Fig. \ref{VortexAntivortex.eps}(a) for the RO case and in \ref{VortexAntivortex.eps}(b) for the ERD case. The parameters used in this figure are $E_{\rm{B}}/E_{\rm{F}}=0.01$ and $h_z/E_{\rm{F}}=0.2$. We chose these parameters to enhance visualization, as the ratio of $\rho_{yy}/\rho_{xx}$ is larger for smaller binding energy. \newline \indent The emergence of elliptic vortices and the structure of the VA pairs in a 2D Fermi superfluid constitute important signatures for the experimentally relevant ERD case. These signatures could be detected during a time-of-flight expansion of the trapped system, or via Bragg spectroscopy, which is also sensitive to the direction of rotation of the supercurrents \cite{phase fluctuations 1}. \newline\indent Having discussed the emergence of vortex-antivortex pairs, we investigate next the BKT transition temperature, where vortex-antivortex unbinding occurs.
\subsection{Berezinskii-Kosterlitz-Thouless critical temperature}\label{Section III G}
\begin{figure*}[tb]
\centering
\includegraphics[width=172mm]{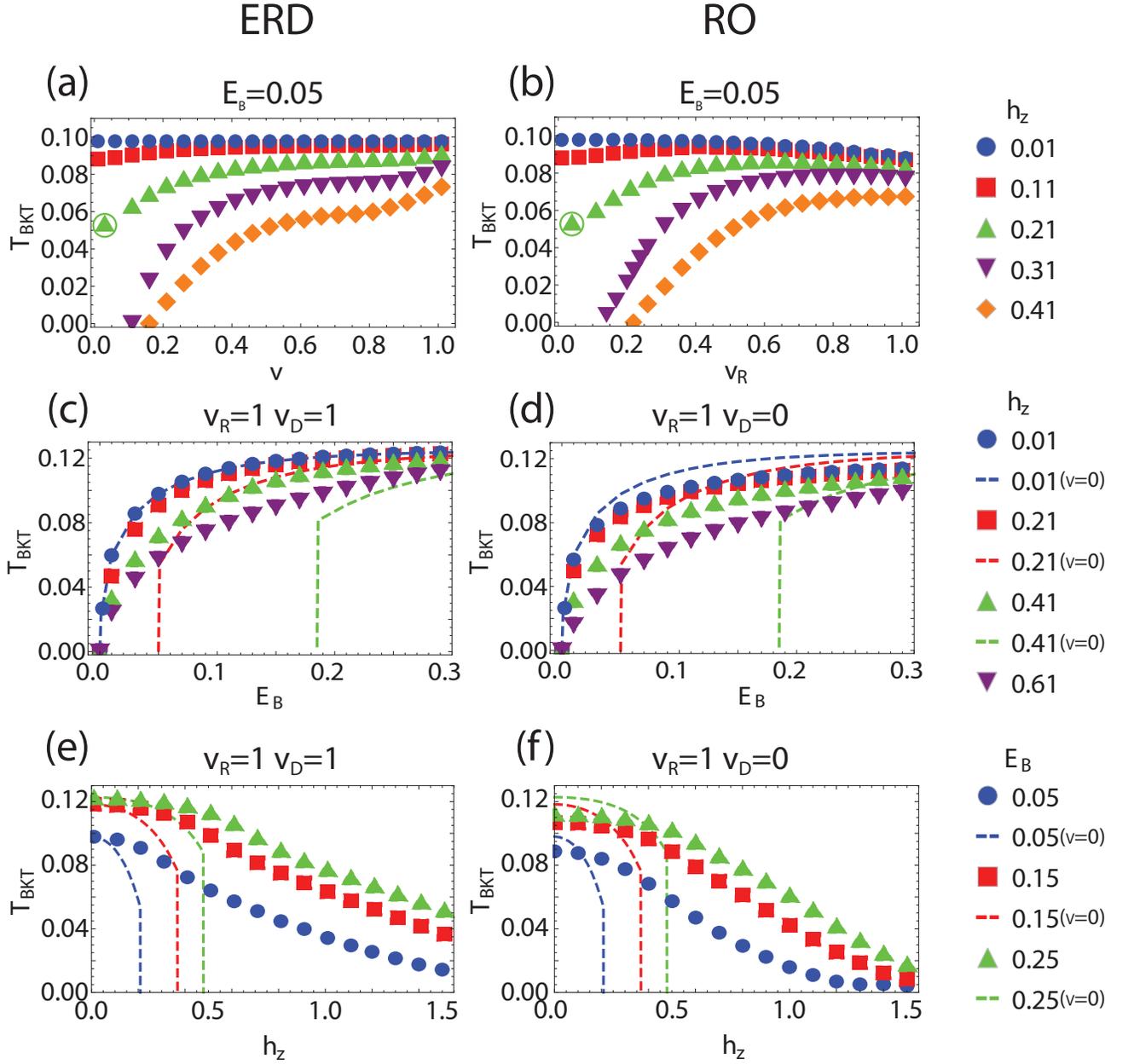}
\caption{Berezinskii-Kosterlitz-Thouless critical temperature, for both the equal-Rashba-Dresselhaus (ERD) case and the Rashba-only (RO) case, shown in the left-hand and right-hand column, respectively. We have plotted $T_{\rm{BKT}}$ as a function of the three main system parameters: the spin-orbit coupling strength (a) and (b), the two-body binding energy (c) and (d) and the Zeeman field (e) and (f). All points shown indicate the BKT critical temperature, except for the two encircled points in (a) and (b), which indicates the mean-field critical temperature $T_{\rm{MF}}$.}
\label{TBKT_6_figures.eps}
\end{figure*}

In this section, we turn to the analysis of the Berezinskii-Kosterlitz-Thouless critical temperature ($T_{\rm{BKT}}$). In order to determine this transition temperature, three equations need to be solved self-consistently: (1) the order parameter equation, determined by the condition $\partial\Omega_{\rm{sp}}/\partial|\Delta|=0$, (2) the number equation $-\partial\Omega_{\rm{sp}}/\partial\mu=n$, and (3) the generalized Kosterlitz-Thouless condition \cite{Nelson-Kosterlitz} $T_{\rm{BKT}}=\frac{\pi}{2}\rho_{\rm{s}}(T_{\rm{BKT}})$, where $\rho_{\rm{s}}=\sqrt{\rho_{xx}\rho_{yy}}$. The first two equations define the saddle-point (mean-field) `transition' temperature $T_{\rm{MF}}$, at which the system changes from its normal phase with $|\Delta|=0$ to its paired phase with $|\Delta|\neq 0$. However, the transition to a true superfluid phase (quasi-condensate) occurs at $T_{\rm{BKT}}<T_{\rm{MF}}$, which is greatly affected by phase fluctuations. Determining this critical temperature requires the simultaneous solution of all three aforementioned equations. \newline \indent Here we note that although the equation $T_{\rm{BKT}}=\frac{\pi}{2}\rho_{\rm{s}}(T_{\rm{BKT}})$ was derived originally in the framework of the phase-only XY-model, it is still justified to use it in the present case with spin-orbit coupling. The reason being that the form of the phase-only action with spin-orbit coupling (\ref{fluctuation action}), can be re-scaled to the phase-only action without spin-orbit coupling (\ref{fluctuation action rescaled}). All the effects of spin-orbit coupling are then contained in the ‘effective’ superfluid density $\rho_{\rm{s}}=\sqrt{\rho_{\rm{xx}}\rho_{\rm{yy}}}$. Because the resulting action is still of the same form as the action for the XY-model, the system with and without spin-orbit coupling fall within the same universality class, when the phase transition is continuous. As such, a simple renormalization group analysis leads to the relation $T_{\rm{BKT}}= \frac{\pi}{2} \rho_{\rm{s}}(T_{\rm{BKT}})$, where $\rho_{\rm{s}}$ is the effective superfluid density, defined above. \newline \indent In Fig. \ref{TBKT_6_figures.eps}, the BKT temperature is shown as a function of the three main system parameters: the spin-orbit coupling strength in (a) and (b), the binding energy in (c) and (d) and the Zeeman field in (e) and (f). Both the equal-Rashba-Dresselhaus (ERD) case and the Rashba-only (RO) case are shown, in the left and right column, respectively. The parameters $v_{\rm{R}},v_{\rm{D}},E_{\rm{B}},h_z$ are the main energy scales of the system, and as such their relative magnitude will significantly affect $T_{\rm{BKT}}$, as is discussed in detail below.\newline \indent
We begin by looking at the effect of the spin-orbit coupling strength on the critical temperature. In Fig. (\ref{TBKT_6_figures.eps})(a) and (b), $T_{\rm{BKT}}$ is shown as a function of the spin-orbit coupling strength in the ERD case ($v_{\rm{R}}=v_{\rm{D}}=v$) and the RO case, respectively, for several values of the Zeeman field $h_z$. Let us first consider the case without Zeeman field (blue filled circles). In this situation, the spin-orbit strength has no influence on the critical temperature in the ERD case. By contrast, in the RO case, the critical temperature is lowered with increasing spin-orbit coupling. The origin of this difference lies in the nature of the spin-orbit coupling itself. In both the ERD and the RO case, the spin-orbit Hamiltonian can be seen as introducing a gauge field, which leads to orbital motion of the fermions. This makes pairing with opposite momenta harder (orbital frustration), thus suppressing superfluidity.
However, in the ERD case, this gauge field can be removed by a gauge transformation, provided that $h_z=0$. \newline \indent Subsequently, we consider the case of non-zero Zeeman field. In the absence of spin-orbit coupling, it is known that Zeeman fields suppress singlet pairing and thus superfluidity. As shown in Fig.(\ref{TBKT_6_figures.eps})(a) and (b), the introduction of spin-orbit coupling increases the critical temperature compared to the case $v_{\rm{R}}=v_{\rm{D}}=0$. The reason for this is that a triplet pairing component is introduced by the presence of spin-orbit coupling, as was demonstrated in section \ref{Section II C} where the generalized helicity basis was discussed. This triplet component cannot be suppressed by the Zeeman field. \newline\indent In the ERD case, $T_{\rm{BKT}}$ rises monotonically and converges to its limiting value without spin-orbit coupling and Zeeman field in the limit $v\rightarrow\infty$. This occurs because in the limit $v\rightarrow\infty$, $h_z$ becomes negligible and the effect of ERD spin-orbit coupling can be gauged away. In the RO case, however, the situation is more complex. For $h_z=0$, the maximum lies at $v_{\rm{R}}=0$. For low values of $h_z$, $T_{\rm{BKT}}$ rises for small values of $v_{\rm{R}}$, reaches a maximum, and then decreases for increasing values of $v_{\rm{R}}$. For high values of $h_z$, $T_{\rm{BKT}}$ is a monotonically increasing function of $v_{\rm{R}}$ that never exceeds the limiting value of $T_{\rm{BKT}}$ at $v_{\rm{R}}\rightarrow\infty$. \newline \indent In Fig. \ref{TBKT_6_figures.eps}(a) and (b) we have encircled two points. These points indicate that the nature of the critical temperature is different there. All other points indicate $T_{\rm{BKT}}$, meaning that the generalized Kosterlitz-Thouless condition $T_{\rm{BKT}}-\pi \rho_{\rm{s}}/2$ is positive for $T>T_{\rm{BKT}}$ and negative for $T<T_{\rm{BKT}}$. However, at these special locations, the situation is different: $T_{\rm{BKT}}-\pi \rho_{\rm{s}}/2$ is still negative for temperatures below this point, but at higher temperatures, the order parameter amplitude $|\Delta|$ is zero, meaning that the mean-field temperature $T_{\rm{MF}}$ has been reached.\newline \indent
In Fig. \ref{TBKT_6_figures.eps} (c) and (d) the Berezinskii-Kosterlitz-Thouless critical temperature is shown as a function of the binding energy $E_{\rm{B}}$, for different values of the Zeeman field $h_z$, for the ERD and the RO cases, respectively. In both figures, the cases with spin-orbit coupling are indicated by symbols, whereas the cases without spin-orbit coupling are indicated by dashed lines. Notice that $T_{\rm{BKT}}$ increases monotonically with increasing binding energy, regardless of the value of the spin-orbit coupling strength. This is as expected, since a larger binding energy leads to more strongly bound fermion pairs, thus strengthening superfluidity. In the asymptotic limit of infinite binding energy, $T_{\rm{BKT}}$ converges to its limiting value: $T_{\rm{BKT}}/T_{\rm{F}}=0.125$, regardless of the strength of the Zeeman field or spin-orbit coupling. This limiting value can be derived from the generalized Kosterlitz-Thouless condition: $T_{\rm{BKT}}=\pi \rho_{\rm{s}}/2$. The superfluid density can at most reach half the total density, where the latter in 2D and in units $\hbar=2m=k_{\rm{F}}=1$ is given by $1/2\pi$. This then leads to $T_{\rm{BKT}}\leq 1/8$. However, when the pairs become bosonic in nature with pair size $\xi_{\rm{pair}}$ much smaller than the interparticle spacing $k_{\rm{F}}^{-1}$, logarithmic corrections due to boson-boson interactions may actually reduce $T_{\rm{BKT}}$, asymptotically \cite{Dalibard,Fisher Hohenberg}.\newline \indent
We deliberately chose not to include these logarithmic corrections to the BKT-critical temperature in the present treatment. Including these corrections involves a more complicated calculation, which lies beyond the scope of the present paper.
It is important, however, to make sure that we do not reach the regime where logarithmic corrections are relevant. To this end, we have calculated the pair size for the case without spin-orbit coupling (Eq. (19) in \cite{Duncan Sa de Melo}) and compared it with the average interparticle distance, given by $k_F^{-1}$. In our paper, the latter quantity equals 1 because of our choice of units. In our case, without spin-orbit coupling, for $h_z/E_F =\{0.01,0.21,0.41\}$ and $E_B/E_F$ varying between 0.03 and 0.35, the pair size varies between 3.32 and 0.84. Thus, at these values of the binding energy, the regime where $\xi_{\rm{pair}} << k_F^{-1}$ has not been reached, validating our approach.
\newline \indent
In the ERD case, we see again that without a Zeeman field, the situations with and without spin-orbit coupling (blue filled circles and blue dashed line, respectively) are exactly equal. When $h_z$ becomes non-zero, however, the effect of spin-orbit coupling on the critical temperature is clear, as $T_{\rm{BKT}}$ is significantly increased. The effect is greater where $h_z$ is larger. Notice also that while $T_{\rm{BKT}}$ is a smooth curve in the presence of spin-orbit coupling, a jump occurs in the case without spin-orbit coupling, due to a first order phase transition. \newline \indent The RO case, shown in Fig. \ref{TBKT_6_figures.eps}(d) is qualitatively similar to the ERD case. The main difference is that in the ERD case, at fixed $h_z$, spin-orbit coupling increases $T_{\rm{BKT}}$ for all values of the binding energy compared to the case without spin-orbit coupling. By contrast, in the RO case, this increase with respect to zero spin-orbit coupling occurs only at low binding energy, while at large binding energy $T_{\rm{BKT}}$ decreases, because of the aforementioned orbital frustration. The binding energy at which $T_{\rm{BKT}}$ crosses the critical temperature without spin-orbit coupling becomes larger with increasing Zeeman field.\newline \indent
Finally, we study the critical temperature as a function of the Zeeman field $h_z$, shown in Fig. \ref{TBKT_6_figures.eps}(e) and (f). The main point of these two figures is the fact that the Clogston limit (i.e. the critical Zeeman field at which the order parameter becomes zero) \cite{Clogston/Chandrasekhar} becomes infinite: the BKT critical temperature only becomes zero in the limit $h_z \rightarrow \infty$. This finding is consistent with the zero-temperature behavior, as discussed in section \ref{Section II E}. There, it was shown that the US-1 phase survives all finite values of the Zeeman field. Because our fluctuation theory only describes the 2D system at non-zero temperature it is comforting that the $T\rightarrow 0$ limit is recovered. In the following section, we connect the zero-temperature phase diagram and its quantum phase transition lines to the Berezinskii-Kosterlitz-Thouless critical temperature.

\subsection{Summarizing 3D phase diagram}\label{Section III H}
\begin{figure*}[tb]
\centering
\includegraphics[width=129mm]{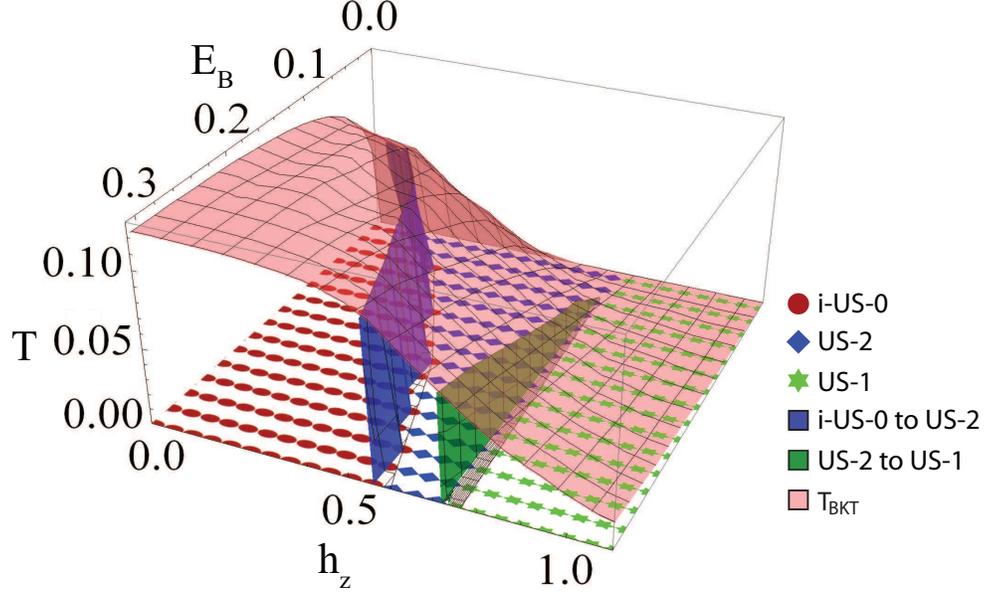}
\caption{Three-dimensional phase diagram, as a function of the two-body binding energy $E_{\rm{B}}$, Zeeman field $h_z$ and temperature $T$, for $v_{\rm{R}}/\tilde{v}_{\rm{F}}=v_{\rm{D}}/\tilde{v}_{\rm{F}}=0.5$. This phase diagram connects the zero-temperature phase diagram to the Berezinskii-Kosterlitz-Thouless critical temperature (red translucent surface). The zero-temperature phase diagram contains three uniform superfluid phases: i-US-0 (red circles), US-2 (blue diamonds) and US-1 (green six-point stars). Finally, we have plotted the boundaries between the different US phases at non-zero temperature: i-US-0 $\rightarrow$ US-2 (blue surface) and US-2 $\rightarrow$ US-1 (green surface).}
\label{3DPhaseDiagram.eps}
\end{figure*}
In Fig. \ref{3DPhaseDiagram.eps}, we show the finite temperature three-dimensional phase diagram as a function of the two-body binding energy and Zeeman field. This figure connects the zero-temperature phase diagram to the Berezinskii-Kosterlitz-Thouless critical temperature, in the ERD case. The zero-temperature phase diagram is plotted in the $(E_{\rm{B}},h_z)$-plane. This phase diagram contains three uniform superfluid phases: i-US-0 (red circles), US-2 (blue diamonds) and US-1 (green six-point stars). The BKT critical temperature as a function of $E_{\rm{B}}$ and $h_z$ is given by the red translucent surface. This surface only reaches $T=0$ in the limit $h_z \rightarrow \infty$ (for all values of $E_{\rm{B}}$), as mentioned in the previous section. \newline \indent Finally, we have extended the quantum phase transition lines to non-zero temperature. The `transition' between i-US-0 and US-2 is given by the blue surface, and is defined as the finite temperature locus where the quasiparticle nodes of the US-2 phase merge at non-zero momentum, leading to the i-US-0 phase. We find that this `transition' has a small temperature dependence: between $T=0$ and $T_{\rm{BKT}}$ this transition line is displaced by an amount on the order of $0.1 h_z$. Similarly, the `transition' between US-2 and US-1 is given by the green surface, which has a negligible dependence on temperature. \newline \indent To the best of our knowledge, no accurate theory exists to calculate the behavior of these lines at non-zero temperature. Here, we obtain these lines based on the quasiparticle description at finite temperatures. This can be regarded as an extrapolation between the zero-temperature results and the Berezinskii-Kosterlitz-Thouless temperature, both of which are known to be qualitatively accurate. Therefore, in the absence of a full theory, it is difficult to claim that these `transition' lines are truly thermodynamic phase boundaries between topologically distinct phases, rather than crossover lines ending at topological phase transition lines at $T=0$. We call them `transition' lines because finite temperature topological invariants can be defined on either side of these lines, provided that a quasiparticle component is still present.
\section{Summary and conclusions}\label{IV}
In this work, we have investigated the effects of spin-orbit coupling on both the zero-temperature and non-zero temperature behavior of a 2D Fermi gas with attractive interactions. We used a generic combination of Rashba and Dresselhaus terms, which allowed us to study both the equal-Rashba-Dresselhaus (ERD) and the Rashba-only (RO) limit. \newline \indent In the first part of the paper, we focused on results at the saddle-point level. Starting from the partition function, we derived the thermodynamic potential within the saddle-point approximation. By minimizing this thermodynamic quantity, the zero-temperature phase diagram was obtained, as a function of the two-body binding energy and Zeeman field. In the ERD case, we identified several topologically distinct uniform superfluid (US) phases, classified according to the nodal structure of the quasiparticle energy bands. We distinguished between the uniform superfluid phase with zero, one or two pairs of nodes (US-0, US-1 and US-2 phases). We found that at any non-zero value of the spin-orbit coupling strength, the system is always in a US phase. More specifically, the US-1 phase survives at any large finite value of the Zeeman field. We identified this behavior by making a momentum dependent transformation to the helicity basis, which diagonalizes the non-interacting Hamiltonian. In this basis, we showed that the order parameter acquires a triplet pairing component, which cannot be suppressed by a Zeeman field. \newline \indent
In the second part of the paper, we focused on fluctuations around the saddle point. By expanding the action up to second order in the phase, the total thermodynamic potential was written as a saddle-point and a fluctuation contribution.  The latter contribution was rescaled to the corresponding action without spin-orbit coupling. We found that the superfluid density becomes anisotropic, due to the presence of spin-orbit coupling (except in the isotropic RO limit). We showed further that the anisotropic sound velocities are sensitive to the quantum phase transition between the US-2 and US-1 phases, and that vortices and antivortices become elliptical instead of circular. Subsequently, we studied the Berezinskii-Kosterlitz-Thouless critical temperature ($T_{\rm{BKT}}$), by simultaneously minimizing the free energy, solving the number equation and satisfying the generalized Kosterlitz-Thouless condition. Our three main findings were as follows. (1) Without the presence of a Zeeman field, ERD spin-orbit coupling can be removed by a gauge transformation, hence $T_{\rm{BKT}}$ remains unchanged compared to the case without spin-orbit coupling. In the RO case, however, increasing the Rashba coupling strength in the absence of a Zeeman field decreases the critical temperature because this introduces orbital frustration for the pairing fermions. (2) In the ERD case, at fixed non-zero Zeeman field, $T_{\rm{BKT}}$ increases relative to the case without spin-orbit coupling for all values of the binding energy. This is due to the emergence of a triplet component of the order parameter, induced by the presence of spin-orbit coupling. However, $T_{\rm{BKT}}$ never becomes larger than the case of vanishing Zeeman and spin-orbit coupling fields, because of residual orbital effects. (3) The Clogston limit becomes infinite when spin-orbit coupling is present, in both the ERD case and the RO case. \newline \indent Finally, we constructed a 3D phase diagram, as a function of the two-body binding energy, Zeeman field and temperature. We have extended the quantum phase transition lines to non-zero temperature, using the nodal structure of the quasiparticle excitation spectrum. The resulting phase diagram connects the zero-temperature result to the BKT-critical temperature, summarizing this paper.

\acknowledgements{One of us (JPAD) wishes to thank E. Vermeyen and D. Sels for interesting and stimulating discussions. JPAD and JT are grateful to N. Verhelst, G. Lombardi and S. Klimin for insightful discussions.  JPAD gratefully acknowledges a post-doctoral fellowship of the Research Foundation-Flanders (FWO-Vlaanderen). This project was supported by projects G.0119.12N, G.0122.12N, G.0429.15N, WOG (WO.035.04N) and by the Research Fund of the University of Antwerp (JT) and ARO (W911NF-09-1-0220) (CARSdM).}\newline


\appendix
\begin{widetext}
\section{Expanding the action up to quadratic order in the phase}\label{Appendix: expanding the action}
In this section, we present the quadratic expansion of the effective action in more detail. As a first step, the second term in Eq. (\ref{Tracelog}) is expanded as

\begin{equation}\label{Expanding the trace}
{\rm Tr }\{\ln[\mathbb{I}+\mathbb{A}^{-1}_{\textbf{k},\omega_n}(0,0) \mathbb{F}_{\textbf{k}}(\theta,\partial\theta)]\}\approx{\rm Tr }[\mathbb{A}^{-1}_{\textbf{k},\omega_n}(0,0)\mathbb{F}_{\textbf{k}}(\theta,\partial\theta)]
-\frac{1}{2}{\rm Tr }[\mathbb{A}^{-1}_{\textbf{k},\omega_n}(0,0)\mathbb{F}_{\textbf{k}}(\theta,\partial\theta)\mathbb{A}^{-1}_{\textbf{k},\omega_n}(0,0)\mathbb{F}_{\textbf{k}}(\theta,\partial\theta)],
\end{equation}
leading to linear and quadratic terms in the expansion, which are treated separately. To calculate both these terms, we require the inverse of the matrix $\mathbb{A}_{\textbf{k},\omega_n}(0,0)$. Using symmetry relations, this inverse matrix can be written as a function of only six elements
\begin{equation}\label{inverse matrix A simplified}
\mathbb{A}^{-1}_{\textbf{k},\omega_n}(0,0)=\frac{1}{D(\textbf{k},\omega_{n})}
\begin{pmatrix}
A_{1,1}(\textbf{k},\omega_{n})&&A_{1,2}(\textbf{k},\omega_{n})&&A_{1,3}(\textbf{k})&&A_{1,4}(\textbf{k},\omega_{n})\\
A_{1,2}^*(\textbf{k},-\omega_{n})&&A_{2,2}(\textbf{k},\omega_{n})&&-A_{1,4}^*(\textbf{k},\omega_{n})&&A_{2,4}(\textbf{k})\\
A_{1,3}^*(\textbf{k})&&-A_{1,4}^*(\textbf{k},\omega_{n})&&-A_{1,1}^*(\textbf{k},\omega_{n})&&A^*_{1,2}(\textbf{k},\omega_{n})\\
A_{1,4}(\textbf{k},\omega_{n})&&A^*_{2,4}(\textbf{k})&&A_{1,2}(\textbf{k},-\omega_{n})&&-A_{2,2}^*(\textbf{k},\omega_{n})
\end{pmatrix}.
\end{equation}
In this expression, the diagonal elements are equal to
\begin{equation}
\begin{array}{l}
A_{1,1}(\textbf{k},\omega_{n})=(-i\omega_{n}+\xi_{\textbf{k}}+h_z)(-i\omega_{n}-\xi_{\textbf{k}}+h_z)(-i\omega_{n}-\xi_{\textbf{k}}-h_z)-(-i\omega_{n}+\xi_{\textbf{k}}+h_z)|h_{\bot}(\textbf{k})|^2-|\Delta|^2(-i\omega_{n}-\xi_{\textbf{k}}-h_z),\\
A_{2,2}(\textbf{k},\omega_{n})=(-i\omega_{n}+\xi_{\textbf{k}}-h_z)(-i\omega_{n}-\xi_{\textbf{k}}+h_z)(-i\omega_{n}-\xi_{\textbf{k}}-h_z)-(-i\omega_{n}+\xi_{\textbf{k}}-h_z)|h_{\bot}(\textbf{k})|^2-|\Delta|^2(-i\omega_{n}-\xi_{\textbf{k}}+h_z).\\
\end{array}
\end{equation}
Furthermore, the non-diagonal elements can be divided into those that depend both on the momentum $\textbf{k}$ and the fermionic Matsubara frequency $\omega_n$
\begin{equation}
\begin{array}{l}
A_{1,2}(\textbf{k},\omega_{n})=-h^*_\bot(\textbf{k})|h_\bot(\textbf{k})|^2-|\Delta|^2h^*_\bot(\textbf{k})+h^*_\bot(\textbf{k})(-i\omega_{n}-\xi_{\textbf{k}}+h_z)(-i\omega_{n}-\xi_{\textbf{k}}-h_z),\\
A_{1,4}(\textbf{k},\omega_{n})=|\Delta|^3+|h_\bot(\textbf{k})|^2|\Delta|-|\Delta|(-i\omega_{n}+\xi_{\textbf{k}}+h_z)(-i\omega_{n}-\xi_{\textbf{k}}+h_z),\\
\end{array}
\end{equation}
and those that only depend on the momentum $\textbf{k}$
\begin{equation}
\begin{array}{l}
A_{1,3}(\textbf{k})=-2h^*_\bot(\textbf{k})|\Delta|(\xi_{\textbf{k}}+h_z),\\
A_{2,4}(\textbf{k})=2h_\bot(\textbf{k})|\Delta|(\xi_{\textbf{k}}-h_z).\\
\end{array}
\end{equation}
Finally, the determinant of the matrix $\mathbb{A}_{\textbf{k},\omega_n}(0,0)$ is denoted by 
\begin{equation}
D(\textbf{k},\omega_{n})=\left(-i\omega_{n}+\epsilon_p^{(+)}(\textbf{k})\right)\left(-i\omega_{n}+\epsilon_p^{(-)}(\textbf{k})\right)\left(-i\omega_{n}-\epsilon_p^{(+)}(\textbf{k})\right)\left(-i\omega_{n}-\epsilon_p^{(-)}(\textbf{k})\right).
\end{equation}
Here, $\epsilon_p^{(\pm)}(\textbf{k})$ denotes the quasiparticle energies. Through the rest of this derivation, we make use of the following symmetry properties of $D(\textbf{k},\omega_n)$:
\begin{equation}\label{symmetries Determinant}
D(\textbf{k},\omega_{n})=D^*(\textbf{k},\omega_{n})=D(\textbf{k},-\omega_{n})=D(-\textbf{k},\omega_{n}).
\end{equation}
We introduce the matrix $\mathbb{B}_{\textbf{k},\omega_n}(\theta,\partial\theta)=\mathbb{A}^{-1}_{\textbf{k},\omega_n}(0,0)\mathbb{F}_{\textbf{k}}(\theta,\partial\theta)$ and then we write the linear term as
\begin{equation}\label{Trace over linear term}
{\rm Tr}[\mathbb{B}_{\textbf{k},\omega_n}(\theta,\partial\theta)]=\frac{1}{\beta L^2}\int d\tau\int d\textbf{r}\sum_\textbf{k}\sum_{\omega_{n}}\frac{1}{D(\textbf{k},\omega_{n})}\bigg(\left[A_{1,1}(\textbf{k},\omega_{n})+A_{2,2}(\textbf{k},\omega_{n})\right]\left[\varepsilon^\theta_+(\textbf{k})-\varepsilon^\theta_-(\textbf{k})\right]\bigg),
\end{equation}
where we used the definition $\varepsilon^\theta_{\pm}(\textbf{k})=\pm \xi^\theta - \zeta_\textbf{k}^{\theta}$, with $\xi^\theta=\frac{i}{2}\frac{\partial\theta_{\textbf{r},\tau}}{\partial\tau}+\frac{1}{4}[\nabla_\textbf{r}(\theta_{\textbf{r},\tau})]^2$ and $\zeta_\textbf{k}^\theta=-\nabla_\textbf{r}(\theta_{\textbf{r},\tau})\cdot\textbf{k}$.
This expression can be simplified by using boundary conditions in calculating the integrals over space and imaginary time. More specifically, we have
\begin{equation}
\int d\tau \dfrac{\partial\theta_{\textbf{r},\tau}}{\partial\tau}=0.
\end{equation}
Furthermore, in (\ref{Trace over linear term}), we have used the fact that $A_{i,i}^*(\textbf{k},\omega_{n})=A_{i,i}(\textbf{k},-\omega_{n})$ with $i=\{1,2\}$. By applying these properties, the linear expansion term can be simplified to:
\begin{equation}\label{Linear expansion term}
{\rm Tr}[\mathbb{B}_{\textbf{k},\omega_n}(\theta,\partial\theta)]=\frac{1}{2\beta L^2}\int d\tau\int d\textbf{r}\sum_\textbf{k}\sum_{\omega_{n}}\widetilde{\mathcal{J}}(\textbf{k},\omega_{n})[\nabla_\textbf{r}(\theta_{\textbf{r},\tau})]^2,
\end{equation}
where we have introduced
\begin{equation}
\widetilde{\mathcal{J}}(\textbf{k},\omega_{n})=\frac{1}{D(\textbf{k},\omega_{n})}\Big(A_{1,1}(\textbf{k},\omega_{n})+A_{2,2}(\textbf{k},\omega_{n})\Big).
\end{equation}
Now, we take a look at the second contribution in the expansion (\ref{Expanding the trace}), which becomes
\begin{equation}\label{A}
-\frac{1}{2}{\rm Tr}[\mathbb{B}_{\textbf{k},\omega_n}(\theta,\partial\theta)\mathbb{B}_{\textbf{k},\omega_n}(\theta,\partial\theta)]=-\frac{1}{2\beta L^2}\int d\tau\int d\textbf{r}\sum_\textbf{k}\sum_{\omega_{n}}\frac{1}{[D(\textbf{k},\omega_{n})]^2}\left(I^{(\varepsilon)}(\textbf{k},\omega_n)+I^{(h)}(\textbf{k},\omega_n)+I^{(\varepsilon h)}(\textbf{k},\omega_n) \right).
\end{equation}
We have divided the integrand in Eq. (\ref{A}) in three main terms, in order to keep track of this extensive expression. The terms in $I^{(\varepsilon)}(\textbf{k},\omega_n)$ do not depend on spin-orbit coupling and are proportional to $[\varepsilon^\theta_+(\textbf{k})]^2$, $[\varepsilon^\theta_-(\textbf{k})]^2$ or $\varepsilon^\theta_+(\textbf{k})\varepsilon^\theta_-(\textbf{k})$:
\begin{equation}\label{I_epsilon}
\begin{aligned}
I^{(\varepsilon)}(\textbf{k},\omega_n)&=\left[A_{1,1}^2(\textbf{k},\omega_n) +A_{2,2}^2(\textbf{k},\omega_n)+2A_{1,2}(\textbf{k},\omega_n)A_{1,2}^*(\textbf{k},-\omega_n)\right]\left\{[\varepsilon^\theta_+(\textbf{k})]^2+[\varepsilon^\theta_-(\textbf{k})]^2\right\}\\
&+ 2\left[|A_{1,3}(\textbf{k})|^2+|A_{2,4}(\textbf{k})|^2+2[A_{1,4}(\textbf{k},\omega_n)]^2 \right]\varepsilon^\theta_+(\textbf{k})\varepsilon^\theta_-(\textbf{k})
\end{aligned}.
\end{equation}
Furthermore, the terms in $I^{(h)}(\textbf{k},\omega_n)$ are proportional to  $(h_{\bot}^\theta)^2$, $[(h_{\bot}^\theta)^*]^2$ or $|h_{\bot}^\theta|^2$, with $h_{\bot}^\theta=-\gamma\frac{\partial\theta_{\textbf{r},\tau}}{\partial y}-i\alpha\frac{\partial\theta_{\textbf{r},\tau}}{\partial x}$:
\begin{equation}\label{I_H}
\begin{aligned}
I^{(h)}(\textbf{k},\omega_n)&=2\left[A_{1,2}^2(\textbf{k},\omega_n)-A_{1,3}(\textbf{k})A_{2,4}^*(\textbf{k}) \right]\left[\left(h_\bot^\theta\right)^*\right]^2+2\left\{\left[A_{1,2}^*(\textbf{k},\omega_n)\right]^2-A^*_{1,3}(\textbf{k})A_{2,4}(\textbf{k}) \right\}\left(h_\bot^\theta\right)^2\\
&+4\left[A_{1,1}(\textbf{k},\omega_n)A_{2,2}(\textbf{k},\omega_n)+|A_{1,4}(\textbf{k},\omega_n)|^2 \right]|h_\bot^\theta|^2
\end{aligned}.
\end{equation}
Finally, the terms in $I^{(\varepsilon h)}(\textbf{k},\omega_n)$ are proportional to the product $\varepsilon^\theta_\pm(\textbf{k})h_{\bot}^\theta$ or $\varepsilon^\theta_\pm(\textbf{k})(h_{\bot}^\theta)^*$:
\begin{equation}\label{I_epsilon*H}
\begin{aligned}
I^{(\varepsilon h)}(\textbf{k},\omega_n)&=2\left[\varepsilon_+^\theta(\textbf{k})+\varepsilon_-^\theta(\textbf{k})\right]\left(h_\bot^\theta\right)^*\\
&\times \left[A_{1,1}(\textbf{k},\omega_n)A_{1,2}(\textbf{k},\omega_n)+A_{1,3}(\textbf{k})A_{1,4}(\textbf{k},\omega_n)-A_{1,2}(\textbf{k},\omega_n)A_{2,2}(\textbf{k},\omega_n)-A_{1,4}(\textbf{k},\omega_n)A_{2,4}^*(\textbf{k}) \right]\\
&+2\left[\varepsilon_+^\theta(\textbf{k})+\varepsilon_-^\theta(\textbf{k})\right]h_\bot^\theta\\
&\times \left[-A_{1,2}^*(\textbf{k},-\omega_n)A_{2,2}(\textbf{k},\omega_n)-A_{1,4}(\textbf{k},\omega_n)A_{2,4}(\textbf{k})-A_{1,1}(\textbf{k},\omega_n)A_{1,2}^*(\textbf{k},-\omega_n)+A_{1,3}^*(\textbf{k})A_{1,4}(\textbf{k},\omega_n) \right]
\end{aligned}.
\end{equation}
At this point, we retain only terms up to quadratic order in the phase field. Moreover, we use the fact that the Matsubara sum runs from $-\infty$ to $\infty$ to group terms together. Expression (\ref{I_epsilon}) then becomes
\begin{equation}
I^{(\varepsilon)}(\textbf{k},\omega_n)=\widetilde{\mathcal{A}}(\textbf{k},\omega_{n})\left(\frac{\partial\theta_{\textbf{r},\tau}}{\partial\tau} \right)^2
+\widetilde{\mathcal{B}}(\textbf{k},\omega_{n})\bigg[\left(\frac{\partial\theta_{\textbf{r},\tau}}{\partial x}\right)^2 k_x^2 + 2\frac{\partial\theta_{\textbf{r},\tau}}{\partial x}\frac{\partial\theta_{\textbf{r},\tau}}{\partial y}k_x k_y + \left(\frac{\partial\theta_{\textbf{r},\tau}}{\partial y}\right)^2 k_y^2\bigg].
\end{equation}
where the following coefficients were defined:
\begin{equation}
\begin{array}{l}
\widetilde{\mathcal{A}}(\textbf{k},\omega_{n})= \frac{1}{2}\left[-A_{1,1}^2(\textbf{k},\omega_{n})+|A_{1,3}(\textbf{k})|^2-A_{2,2}^2(\textbf{k},\omega_{n})+|A_{2,4}(\textbf{k})|^2 -2A_{1,2}(\textbf{k},\omega_{n})A_{1,2}^*(\textbf{k},-\omega_{n})+2A_{1,4}^2(\textbf{k},\omega_{n})\right]\\
\widetilde{\mathcal{B}}(\textbf{k},\omega_{n})=2\left[A_{1,1}^2(\textbf{k},\omega_{n})+|A_{1,3}(\textbf{k})|^2+A_{2,2}^2(\textbf{k},\omega_{n})+|A_{2,4}(\textbf{k})|^2+2A_{1,2}(\textbf{k},\omega_{n})A_{1,2}^*(\textbf{k},-\omega_{n})+2A_{1,4}^2(\textbf{k},\omega_{n})\right]
\end{array}.
\end{equation}
Similarly, expression (\ref{I_H}) can be written as
\begin{equation}
I^{(H)}(\textbf{k},\omega_n)= \widetilde{\mathcal{C}}(\textbf{k},\omega_{n})\left(\frac{\partial\theta_{\textbf{r},\tau}}{\partial x}\right)^2 + \widetilde{\mathcal{D}}(\textbf{k},\omega_{n})\frac{\partial\theta_{\textbf{r},\tau}}{\partial x}\frac{\partial\theta_{\textbf{r},\tau}}{\partial y} + \widetilde{\mathcal{E}}(\textbf{k},\omega_{n})\left(\frac{\partial\theta_{\textbf{r},\tau}}{\partial y}\right)^2,
\end{equation}
with the following coefficients:
\begin{equation}\label{C,D,E}
\begin{array}{l}
\widetilde{\mathcal{C}}(\textbf{k},\omega_{n})=4\alpha^2\left[-A_{1,2}^2(\textbf{k},\omega_{n})+A_{1,3}(\textbf{k})A_{2,4}^*(\textbf{k})+A_{1,1}(\textbf{k},\omega_{n})A_{2,2}(\textbf{k},\omega_{n})+|A_{1,4}(\textbf{k},\omega_{n})|^2 \right]\\
\widetilde{\mathcal{D}}(\textbf{k},\omega_{n})=-4i\alpha\gamma\left[A_{1,2}^2(\textbf{k},\omega_{n})-[A_{1,2}^*(\textbf{k},\omega_{n})]^2-A_{1,3}(\textbf{k})A_{2,4}^*(\textbf{k})+A_{1,3}^*(\textbf{k})A_{2,4}(\textbf{k}) \right]\\
\widetilde{\mathcal{E}}(\textbf{k},\omega_{n})=4\gamma^2\left[A_{1,2}^2(\textbf{k},\omega_{n})-A_{1,3}(\textbf{k})A_{2,4}^*(\textbf{k})+A_{1,1}(\textbf{k},\omega_{n})A_{2,2}(\textbf{k},\omega_{n})+|A_{1,4}(\textbf{k},\omega_{n})|^2 \right]
\end{array}.
\end{equation}
In (\ref{C,D,E}), we have made use of the fact that all terms proportional to $k_xk_y$ vanish, because these are odd under a parity transformation of the integral over momentum $\textbf{k}$. Using the same reasoning, we can see that the momentum integral over $\widetilde{\mathcal{D}}(\textbf{k},\omega_n)$ is equal to zero. Finally, we look at expression (\ref{I_epsilon*H}), which can be written as
\begin{equation}
I^{(\varepsilon H)}(\textbf{k},\omega_n)=\widetilde{\mathcal{F}}(\textbf{k},\omega_{n})\left(\frac{\partial\theta_{\textbf{r},\tau}}{\partial x} \right)^2+\widetilde{\mathcal{G}}(\textbf{k},\omega_{n})\frac{\partial\theta_{\textbf{r},\tau}}{\partial x}\frac{\partial\theta_{\textbf{r},\tau}}{\partial y}
+\widetilde{\mathcal{H}}(\textbf{k},\omega_{n})\left(\frac{\partial\theta_{\textbf{r},\tau}}{\partial y} \right)^2,
\end{equation}
with the following coefficients
\begin{equation}\label{F,G,H}
\begin{array}{l}
\widetilde{\mathcal{F}}(\textbf{k},\omega_{n})=-4i\alpha k_x\Gamma_x(\textbf{k},\omega_{n})\\
\widetilde{\mathcal{G}}(\textbf{k},\omega_{n})=4[\gamma k_x \Gamma_y(\textbf{k},\omega_{n})-i\alpha k_y \Gamma_x(\textbf{k},\omega_{n})]\\
\widetilde{\mathcal{H}}(\textbf{k},\omega_{n})=4\gamma k_y \Gamma_y(\textbf{k},\omega_{n})
\end{array},
\end{equation}
where $\Gamma_x(\textbf{k},\omega_{n})$ is defined by
\begin{equation}
\begin{aligned}
\Gamma_x(\textbf{k},\omega_{n})=&A_{1,1}(\textbf{k},\omega_{n})A_{1,2}(\textbf{k},\omega_{n})-A_{1,3}(\textbf{k})A_{1,4}(\textbf{k},\omega_{n})-A^*_{1,2}(\textbf{k},-\omega_{n})A_{2,2}(\textbf{k},\omega_{n})\\
&-A_{1,4}(\textbf{k},\omega_{n})A_{2,4}(\textbf{k})-A_{1,1}(\textbf{k},\omega_{n})A_{1,2}^*(\textbf{k},-\omega_{n})+A_{1,3}^*(\textbf{k})A_{1,4}(\textbf{k},\omega_{n})\nonumber\\&+A_{1,2}(\textbf{k},\omega_{n})A_{2,2}(\textbf{k},\omega_{n})+A_{1,4}(\textbf{k},\omega_{n})A^*_{2,4}(\textbf{k})\\
\end{aligned},
\end{equation}
and $\Gamma_y(\textbf{k},\omega_{n})$ is defined by
\begin{equation}
\begin{aligned}
\Gamma_y(\textbf{k},\omega_{n})=&A_{1,1}(\textbf{k},\omega_{n})A_{1,2}(\textbf{k},\omega_{n})-A_{1,3}(\textbf{k})A_{1,4}(\textbf{k},\omega_{n})+A^*_{1,2}(\textbf{k},-\omega_{n})A_{2,2}(\textbf{k},\omega_{n})\\
&+A_{1,4}(\textbf{k},\omega_{n})A_{2,4}(\textbf{k})+A_{1,1}(\textbf{k},\omega_{n})A_{1,2}^*(\textbf{k},-\omega_{n})-A_{1,3}^*(\textbf{k})A_{1,4}(\textbf{k},\omega_{n})\nonumber\\&+A_{1,2}(\textbf{k},\omega_{n})A_{2,2}(\textbf{k},\omega_{n})+A_{1,4}(\textbf{k},\omega_{n})A^*_{2,4}(\textbf{k})
\end{aligned}.
\end{equation}
Using elementary algebra, we can show that the former two coefficients can be written as $\Gamma_x(\textbf{k},\omega_{n})\propto  k_x f(k_x^2,k_y^2)$ and $\Gamma_y(\textbf{k},\omega_{n}) \propto k_y f(k_x^2,k_y^2)$, where $f$ is a real function depending only on the square of the momentum components. Applying this form to expression (\ref{F,G,H}), we see that expression $\widetilde{\mathcal{G}}(\textbf{k},\omega_{n})$ is proportional to $k_xk_y$ and that its momentum integral is zero.
Putting everything together, the total action can be written in a compact generic form:
\begin{equation}\label{The fluctuation action}
S_{\rm{fl}}=\frac{1}{2}\int d\tau \int d\textbf{r}\left[\mathcal{A}\left(\dfrac{\partial\theta}{\partial\tau} \right)^2 +
\begin{pmatrix}
\dfrac{\partial\theta}{\partial x}&&\dfrac{\partial\theta}{\partial y}
\end{pmatrix}
\begin{pmatrix}
\rho_{xx} && \rho_{xy}\\
\rho_{yx} && \rho_{yy}
\end{pmatrix}
\begin{pmatrix}
\partial\theta/ \partial x\\\partial\theta/ \partial y
\end{pmatrix}\right].
\end{equation}
Here, we defined the coefficient
\begin{equation}
\mathcal{A}=\frac{1}{2\beta L^2}\sum_{\textbf{k},\omega_{n}}\widetilde{\mathcal{A}}(\textbf{k},\omega_{n}),
\end{equation}
which plays the role of the compressibility of the superfluid,
and
\begin{equation}\label{rho_xx rho_xy rho_yy}
\begin{dcases}
\rho_{xx}=\frac{1}{2\beta L^2}\sum_{\textbf{k},\omega_{n}}\left(\beta+\widetilde{\mathcal{B}}(\textbf{k},\omega_{n})k_x^2+\widetilde{\mathcal{C}}(\textbf{k},\omega_{n})+\widetilde{\mathcal{F}}(\textbf{k},\omega_{n})-\widetilde{\mathcal{J}}(\textbf{k},\omega_{n}) \right)\\
\rho_{xy}=\rho_{yx}=\frac{1}{4\beta L^2}\sum_{\textbf{k},\omega_{n}}\left(2\widetilde{\mathcal{B}}(\textbf{k},\omega_{n})k_x k_y+\widetilde{\mathcal{D}}(\textbf{k},\omega_{n})+\widetilde{\mathcal{G}}(\textbf{k},\omega_{n}) \right)\\
\rho_{yy}=\frac{1}{2\beta L^2}\sum_{\textbf{k},\omega_{n}}\left(\beta+\widetilde{\mathcal{B}}(\textbf{k},\omega_{n})k_y^2+\widetilde{\mathcal{E}}(\textbf{k},\omega_{n})+\widetilde{\mathcal{H}}(\textbf{k},\omega_{n})-\widetilde{\mathcal{J}}(\textbf{k},\omega_{n}) \right)\\
\end{dcases},
\end{equation}
which play the role of the superfluid density tensor components $\rho_{ij}$.
As mentioned previously, both the momentum integral over $\widetilde{\mathcal{D}}(\textbf{k},\omega_n)$ and $\widetilde{\mathcal{G}}(\textbf{k},\omega_n)$ are zero. Furthermore, because all terms in $\widetilde{\mathcal{B}}(\textbf{k},\omega_n)$ are proportional to $k_x^2$ or $k_y^2$, this part of the integral also vanishes due to the factor $k_xk_y$. Hence, we have that $\rho_{xy}=\rho_{yx}=0$, making the superfluid density tensor $\rho_{ij}$ diagonal.\newline
In the limit of spin-orbit coupling tending to zero, the action (\ref{The fluctuation action}) reduces to the known form \cite{phase fluctuations 1}.

\end{widetext}

\end{document}